\documentclass[journal=jpcm,10pt,amsmath,amssymb]{iopart}
\usepackage{graphicx,epsfig}
\usepackage{graphicx,amsfonts}
\usepackage{iopams}
\usepackage{color}
\usepackage{cite}
\usepackage{wasysym}
\usepackage[normalem]{ulem}
\usepackage{bm}
\usepackage{cuted}
\usepackage{graphicx}
\usepackage[colorlinks, allcolors=blue]{hyperref}
\begin{document}
%%%%%%%%%%%%%%%%%%%%%%%%%%%%%%%%%%%%%%%%%%%%%%%%%%%
%                               TITLE & ABSTRACT
%%%%%%%%%%%%%%%%%%%%%%%%%%%%%%%%%%%%%%%%%%%%%%%%%%%
%Title of paper
\title[]{Magnetic plateaus and jumps in a spin-1/2 ladder with alternate Ising-Heisenberg rungs: a field dependent study}\author{Sk Saniur Rahaman$^{1}$}
\author{Manoranjan Kumar$^{1,}$}
\author{Shaon Sahoo$^{2,}$}
\ead{manoranjan.kumar@bose.res.in}
\ead{shaon@iittp.ac.in}
\footnote[3]{The last two authors have equal contributions.}
\address{$^1$ S. N. Bose National Centre for Basic Sciences, Block JD, Sector III, Salt Lake, Kolkata 700106, India}
\address{$^2$ Department of Physics, Indian Institute of Technology, Tirupati, India} 
\date{\today}
\begin{abstract}
We study a frustrated two-leg spin-1/2 ladder with alternate Ising and isotropic Heisenberg rung exchange interactions, whereas, interactions along legs and diagonals are Ising type. The ground-state (GS) of this model has four exotic phases: (i) the stripe rung ferromagnet (SRFM), (ii) the anisotropic anti-ferromagnet (AAFM), (iii) the Dimer, and (iv) the stripe leg ferromagnet (SLFM) in absence of any external magnetic field. In this work, we study the effect of externally applied longitudinal and transverse fields on  GS phases and note that there are two plateaus with per-site magnetization $1/4$ and $1/2$. There is another plateau at zero magnetization due to a finite spin gap in the presence of a longitudinal field. The exact diagonalization (ED) and the transfer matrix (TM) methods are used to solve the model Hamiltonian and the mechanism of plateau formation is analyzed using spin density, quantum fidelity, and quantum concurrence. In the (i) SRFM phase, Ising exchanges are dominant for all spins but the Heisenberg rungs are weak, and therefore, the magnetization shows a continuous transition as a function of the transverse field. In the other three phases [(ii)-(iv)], the Ising dimer rungs are weak and those are broken first to reach a plateau with per-site magnetization $1/4$, having a large gap which is closed by further application of the transverse field. 
\end{abstract}
%\pacs{75.10.Jm, 03.65.Vf}

\maketitle
\ioptwocol
%%%%%%%%%%%%%%%%%%%%%%%%%%%%%%%%%%%%%%%%%%%%%%%%%%%%%%%%%%%%%%%%%%
%																 %
%		INTRODUCTION											 %
%																 %
%%%%%%%%%%%%%%%%%%%%%%%%%%%%%%%%%%%%%%%%%%%%%%%%%%%%%%%%%%%%%%%%%%

%{\it Introduction.}-- 
\section{Introduction}
\label{sec:introduction}
 Frustrated low-dimensional quantum magnets exhibit a diverse range of quantum phases that generate significant interest among both theorists and experimentalists. Consequently, theoretical studies are crucial for verifying experimental results, especially due to the continuous synthesis of low-dimensional magnetic materials\cite{nakamura1998theoretical,guo2020crystal,kakarla2021single,dagotto1996,chubukov1991, chubukov1994theory, hutchings1979, park2007, mourigal2012, drechsler2007, dutton201224, dutton2012108, maeshima2003, sandvik1996, johnston1987}. In the spin chains and ladder systems, the competing exchange interactions lead to many interesting quantum phases like ferromagnetic ground state (GS) \cite{vekua2003phase}, N\'eel phase \cite{richter2004quantum, ogino2021continuous,ogino2022ground}, Luttinger liquid \cite{luttinger1963exactly,sakai2022field}, spiral \cite{dmaiti2019}, spin liquid \cite{balents2010spin, liao2017gapless}, dimer phase \cite{dagotto1996}, etc. The antiferromagnetic isotropic Heisenberg spin-1/2 zigzag ladder or the $J_1-J_2$ chain has a gapless spectrum in strong leg coupling limit, whereas, it has a gapped spectrum for the moderate value of the ratio of the exchange interactions due to dimerization along the rung. \cite{bethe1931theorie,barnes1993excitation, johnston2011magnetic, mkumar2015,chitra1995,srwhite1996,rahaman2023machine}. 
 
The anisotropy in the exchange interaction significantly influences the GS properties. For example, the isotropic Heisenberg spin-1/2 $J_1-J_2$ model in the small $J_2$ limit has a gapless spectrum and its GS is a spin liquid phase with quasi-long range order (QLRO)\cite{bethe1931theorie,chitra1995,srwhite1996}. But, the spectrum of the XXZ Heisenberg spin-1/2 chain is gapped for a large axial anisotropy $\Delta >1$, and in this limit, the spins behave like Ising type,  whereas, the spectrum is gapless for $\Delta<1$ and spins are aligned in XY plane \cite{PhysRevB.93.054417,mikeska2004one}. The GS of an isotropic Heisenberg  spin-1/2 normal ladder exhibits short range order and the spin gap is finite for any non-zero value of rung exchange. On the other hand, the spectrum of the anisotropic ladder systems can be gapless \cite{wessel2017efficient,morita2015quantum,hijii2005phase,mikeska2004one,liu2022gapless,li2017groundstate,PhysRevB.99.224418}. In a spin-1/2 ladder with isotropic rung and axial anisotropic leg exchange $\Delta$, the GS can be tuned from singlet to N\'eel phase by increasing $\Delta$ \cite{hijii2005phase}. But, for a normal spin-1/2 ladder with anisotropy in both leg and rung exchange interactions, the GS can be XY, N\'eel, or rung singlet (RS) phase on tuning the rung exchange and axial anisotropy \cite{li2017groundstate}. Another type of anisotropic spin-1/2 ladder is the Kitaev-Heisenberg model on a two-leg ladder where the GS has many exotic quantum phases \cite{PhysRevB.99.224418}.

Many spin-1/2 ladders with isotropic and anisotropic exchange interactions are reported to display magnetic plateaus and jumps in the magnetization curve by tuning the external magnetic field \cite{ivanov2009spin, sakai1999critical, sakai2000magnetic,naseri2017magnetic,zad2018phase, japaridze2006magnetization, moradmard2014,dey2020magnetization,PhysRevLett.78.1984,
strecka2003existence,verkholayak2013,karl2019breakdown,verkholayak2012,strecka2014,strevcka2014magnetization}. Japaridze et al. studied a two-leg spin-$\frac{1}{2}$ ladder system with leg interaction $J_{\parallel}$ and alternate rung as $J_{\perp}^+$, $J_{\perp}^-$ in presence of a longitudinal field $h$, and show that the system have zero magnetization plateau up to $h_{c1}^-$, a plateau at half of the saturation magnetization for the magnetic field between $h_{c1}^+$ and $h_{c2}^-$ and fully polarized spin is achieved at $h_{c2}^+$ \cite{japaridze2006magnetization}. Moradmard et al. studied a spin-1/2 ladder system with XXZ interaction and they showed different phases like x-FM, z-FM, y-N\'eel, and z-N\'eel in a magnetic phase diagram in the plane of anisotropy interaction $\Delta$ and magnetic field $h$ \cite{moradmard2014}. Similarly, Dey et.al. carried out the magnetization study of isotropic Heisenberg spin-1/2 on a 5/7-skewed ladder and showed multiple plateau phases with field $h$ \cite{dey2020magnetization}. They also note that plateau phases are the consequences of gaps in the spectrum, and these plateau phases can be explained in terms of Oshikawa, Yamanaka, and Affleck (OYA) criterion \cite{PhysRevLett.78.1984}. 

Some of the real compounds like the hetero-bimetallic coordination polymer [$(Tp)_2Fe_2 (CN)_6$-$(OAc)(bap)$ $Cu_2 (CH_3OH)_2CH_3OH ·
H_2O$] forms an effective Ising-Heisenberg spin-1/2 branched chain model \cite{karl2019breakdown}, whereas,  $(VO)_2P_2O_7$, $CaV_2O_5$ and $MgV_2O_5$ can be modeled by using two legs spin-1/2 ladders with the Ising-Heisenberg exchange \cite{verkholayak2012}. The quantum phase diagram, magnetization curves, and concurrence of the spin-1/2 Ising Heisenberg branched chain are studied using the transfer-matrix method and they note the plateau at 1/2 of saturation magnetization \cite{karl2019breakdown}. For a two-leg spin-$1/2$ ladder with Ising exchange along the leg, diagonal, and Heisenberg exchange along the rung, Verkholayak et. al. noted that on applying a longitudinal magnetic field, the  N\'eel phase of the GS undergoes a phase transition to a half plateau or staggered bond (SB) phase in presence of moderate field and to a fully spin-polarized state for strong field limit \cite{verkholayak2012}. With few similarities to these ladder systems, we propose another type of anisotropic spin-1/2 two-leg ladder which may have various types of plateau at 0, 1/2, and full of saturation magnetization. The quantum phase transition in these systems can alternatively be determined using the concurrence \cite{PhysRevLett.80.2245, karlova2020} and fidelity \cite{PhysRevE.79.031101} calculations.
 
\begin{figure}[t]
\includegraphics[width=1.0\linewidth]{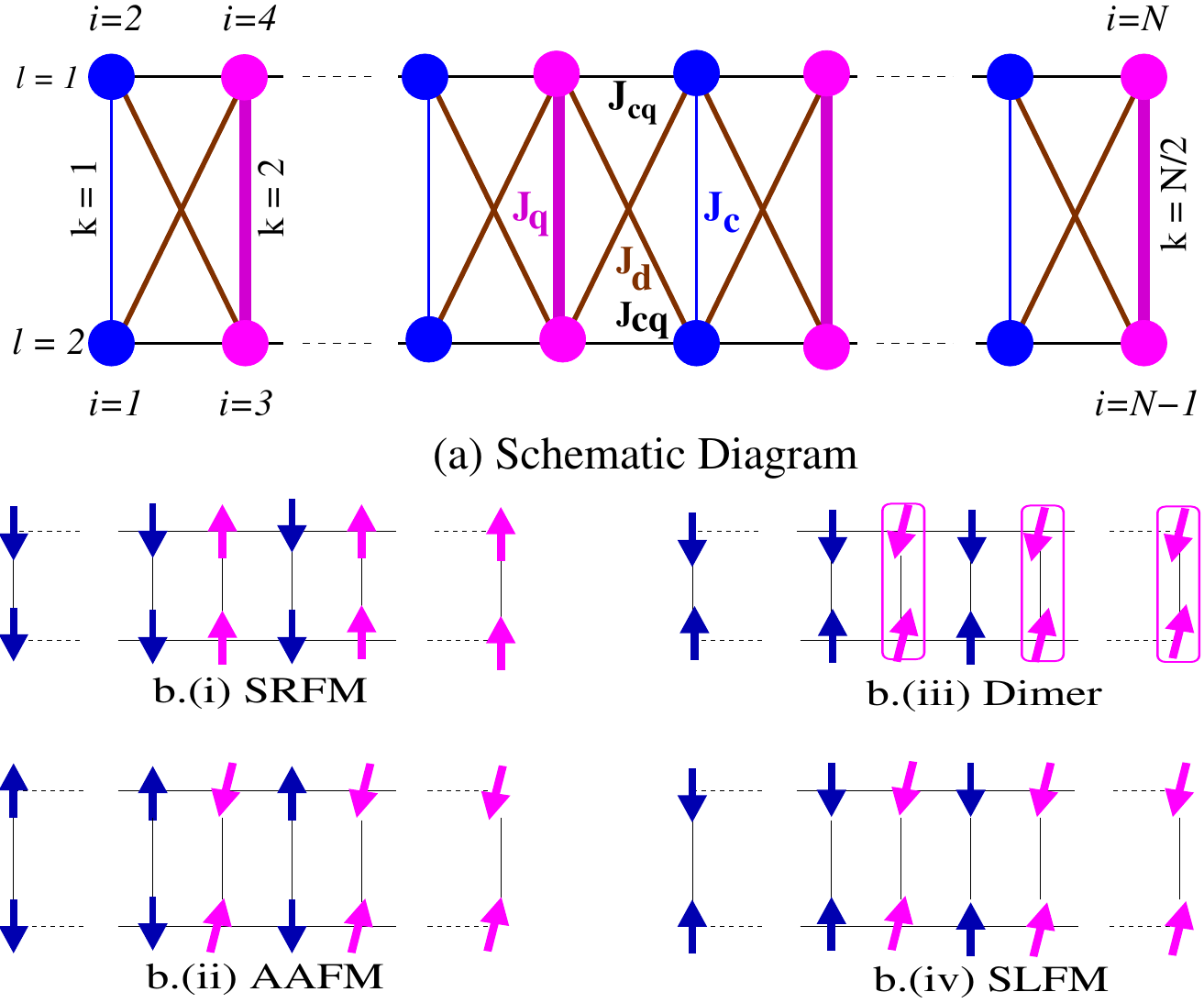}
\caption{(color online) (a) Schematic diagram of the spin ladder with alternate Ising-Heisenberg rung interactions. $J_c$ and $J_q$ are the alternative Ising and Heisenberg type rung interactions respectively. $J_{cq}$, $J_d$ are the Ising type leg and diagonal exchange interactions respectively. Blue and magenta color circles represent $\sigma$ and S spins in Eq.\ref{eqn:hamiltonian} respectively. $l$, $k$, and $i$ represent leg and  rung, and site indices respectively. The spin configurations of four exotic phases with $J_c=J_{cq}=1$: b.(i) SRFM ($J_q=0.2$, $J_d=2.0$), b.(ii) AAFM ($J_q=2.0$, $J_d=0.4$), b.(iii) Dimer ($J_q=2.0$, $J_d=1.0$), and b.(iv) SLFM ($J_q=2.0$, $J_d=1.6$) are shown. Blue and magenta rung pairs are representing $\sigma-\sigma$ and $S-S$ rung pairs. The boxes shown in Subfigure b.(iii) represent perfect singlets. Quantum phases are studied earlier in Ref. \cite{Saniur_Rahaman_2021}.}
\label{fig:model2}
\end{figure}

We consider a spin-1/2  frustrated two-leg ladder system with alternating Ising and Heisenberg type rung exchanges, where the diagonal and leg exchange are of Ising type as shown in Fig.\ref{fig:model2}.a. In this model, $J_c$ and $J_q$ are the alternate Ising and Heisenberg rung exchange interactions respectively, where, $J_{cq}$ and $J_d$ are the leg and diagonal exchange interaction strengths of Ising type respectively. The quantum phase diagram of this model is studied earlier in parameter space of $J_q$ and $J_d$ (both are antiferromagnetic) by considering $J_c = J_{cq} = 1$ \cite{Saniur_Rahaman_2021}. The system exhibits four distinct GS phases under periodic boundary condition (PBC): (i) stripe rung ferromagnet (SRFM), (ii) the anisotropic antiferromagnet (AAFM), (iii) the Dimer, and (iv) the stripe leg ferromagnet (SLFM) depending upon the values of $J_q$ and $J_d$ in absence of any magnetic field \cite{Saniur_Rahaman_2021}. 
The GS phases are schematically represented in Fig\ref{fig:model2}.b.[(i)-(iv)]. The SRFM phase is Ising dominated where all the spins are aligned along the z direction completely. Whereas, the [(ii)-(iv)] phases are anisotropic. Although all the rungs are anti-ferromagnetically aligned for the [(ii)-(iv)] phases, the spin arrangements along the leg are different. Along the leg, the alignment of the spins is anti-ferromagnetic for the AAFM phase and ferromagnetic for the SLFM phase. In the Dimer phase, there is no spin alignment along the leg. In this manuscript, we study the effect of both longitudinal and transverse magnetic fields on the four GS phases with a few sets of $J_{q}$, $J_d$  values discussed in Sec.\ref{sec:results}.

 In this work, we observe that in all the quantum phases, the system exhibits a plateau at 0, 1/2, and 1 of saturation magnetization in the presence of an externally applied longitudinal magnetic field. The calculations are done using the exact diagonalization (ED) \cite{er1975iterativecalculationof} and transfer matrix (TM) \cite{suzuki1985transfer} methods, and results from both methods agree excellently with each other. Furthermore, we calculate the zero-temperature limit quantum fidelity, fidelity susceptibility, and quantum concurrence from the partition function of the ladder using TM, and find that these results are in accordance with the exact calculation. The study of the magnetization under a transverse field is carried out using ED only, and it is noticed that the magnetization shows a half, and full of saturation magnetization plateaus. 

This paper is divided into a few sections as follows. First, the model is discussed briefly in Sec. \ref{sec:modelham}. This is followed by a discussion on methods in Sec. \ref{sec:methods}. In Sec. \ref{sec:magz}, the magnetization process is discussed in the presence of the longitudinal field. In Sec. \ref{sec:quant_fid} 
we discuss the zero-temperature limit quantum fidelity and bipartite concurrence for different phases. In Sec. \ref{sec:magnetic_phase_diagrams}, the quantum phase diagrams are shown for four different longitudinal fields. In Sec. \ref{sec:magx}, we discuss the magnetization process in the presence of a transverse field. 
In Sec. \ref{sec:summary}, we summarise the results and conclude the paper.

\section{Model Hamiltonian}
\label{sec:modelham}
We construct the Hamiltonian for a spin-1/2 two-leg ladder with $N$ number of spins periodically connected along the leg, which turns out to be comprised of $n=\frac{N}{4}$ number of unit cells. In each unit cell, one rung pair is connected through an Ising type exchange $J_c$, whereas, the other one is coupled with a Heisenberg type exchange $J_q$ as shown in Fig.\ref{fig:model2}.(a). These rungs couple each other through Ising type exchanges: $J_{cq}$ along the leg, $J_d$ along the diagonal. The spins with rung coupling $J_c$ and $J_q$ are marked with $\sigma$ and $\vec{S}$ respectively. Here onward, the Ising type and Heisenberg type rung spin pairs are to be called $\sigma-\sigma$ and $S-S$ pairs respectively. Let us now write down the Hamiltonian for the $j^{th}$ unit cell as
\begin{eqnarray} 
\label{eqn:hamiltonian}
 &&  \mathbf{H_j}= J_{q}^{z} S^{z}_{2j,1} S^{z}_{2j,2}
         +\frac{J_{q}^{xy}}{2} \bigg[ S^{+}_{2j,1} S^{-}_{2j,2} + S^{-}_{2j,1} S^{+}_{2j,2}\bigg]
   \nonumber \\
 &&               +\frac{J_c}{2} \bigg[\sigma_{2j-1,1}\sigma_{2j-1,2}
    +\sigma_{2j+1,1}\sigma_{2j+1,2}\bigg]+J_{cq} \times \nonumber \\
 &&   \bigg[ S_{2j,1}^{z}(\sigma_{2j-1,1}+\sigma_{2j+1,1})
    +S_{2j,2}^{z}(\sigma_{2j-1,2}+\sigma_{2j+1,2})\bigg] \nonumber \\
 &&               + J_{d} \times \nonumber \\
 &&   \bigg[ S_{2j,1}^{z} (\sigma_{2j-1,2}+\sigma_{2j+1,2})
    + S_{2j,2}^{z} (\sigma_{2j-1,1}+\sigma_{2j+1,1})\bigg] \nonumber \\ 
 &&               - \frac{h}{2} \sum_{l=1}^2(2S_{2j,l}^{z}+\sigma_{2j-1,l}
    +\sigma_{2j+1,l}) \nonumber \\
 &&               - \frac{h^x}{2} \sum_{l=1}^2(2S_{2j,l}^{x}+\sigma_{2j-1,l}^{x} 
    +\sigma_{2j+1,l}^{x}) 
\end{eqnarray}

Here, $h$ and $h^x$  are the longitudinal (+z direction) and transverse ( +x direction) fields respectively. $S^x$, $S^z$ are the spin components along +x, +z respectively of the spin $\vec{S}$, whereas, $S^{+}$, $S^{-}$ are the raising and lowering operators respectively of the same. In Eq.\ref{eqn:hamiltonian}, we consider $J_c=J_{cq}=1$ and $J_q^{z}=J_q^{xy}=J_q$ throughout our manuscript. The general Hamiltonian of the ladder under PBC with system size $N$ is the summation of the $n$ unit cells, which can be written as $\mathbf{H} = \sum_{j=1}^{n} \mathbf{H_j}$. 
\section{Methods}
\label{sec:methods}
We employ the ED method to solve the energy eigenvalues and eigenvectors of the Hamiltonian in Eq.\ref{eqn:hamiltonian} for the system sizes $N=16, 20, 24$ in the presence of longitudinal and transverse fields both. Whereas, in the absence of a transverse field i.e., for $h^x=0$, the Hamiltonian of two consecutive units commute to each other, and so we employ the TM method to calculate the magnetization, quantum fidelity, quantum concurrence from free energy, and partition function. The partition function for the entire ladder with system size $N$ (or $n=N/4$ unit) can be written as $Q_N(h,\beta)=Tr(e^{-\beta \mathbf{H}})=[Q_4(h,\beta)]^n$ (see Appendix \ref{appendix}). $Q_4(h,\beta)$ is the partition function for one unit of 4 spins and $\beta$ is the inverse temperature. For this model, $Q_N(h,\beta)=[\lambda_1^n+\lambda_2^n+\lambda_3^n+\lambda_4^n]$ , where, $\lambda_1, \lambda_2,\lambda_3, \lambda_4$ are the eigenvalues of a $4 \times 4$ transfer matrix for one unit (see Appendix \ref{appendix}). In the limit $n \rightarrow \infty$, and with the condition $\lambda_1 \gg \lambda_2 \gg \lambda_3 \gg \lambda_4$, one can write $Q_N(h,\beta) \approx \lambda_{1}^n$ and $Q_4(h,\beta) \approx \lambda_{1}$. At zero-temperature limit i.e., for $\beta\rightarrow\infty$, after defining some of the system parameters: $\Delta_2=\sqrt{1+4(\frac{1-J_d}{J_q^{xy}})^2}$, $Q=e^{\beta J_q^{z}/4}$, we obtain the partition function for one unit (from the Eq. \ref{eqn:approx_Q4} in Appendix \ref{appendix})
\begin{eqnarray} 
\label{eqn:approximated}
\centering
& Q_4(h,\beta)= \nonumber \\
& 2e^{\frac{\beta (J_c)}{4}} \times \nonumber \\
& \big[2 Q^{-1}Cosh[\beta h]+Q Cosh[\frac{\beta J_q^{xy}}{2}]+Q Cosh[\frac{\beta J_q^{xy}\Delta_2}{2}]
\big] \nonumber \\
& +2e^{\frac{-\beta (J_c-4h)}{4}} \times \nonumber \\
&   \bigg[
Q^{-1}Cosh[\beta (h-(J_{cq}+J_d))] +Q Cosh[\frac{\beta J_q^{xy}}{2}]
\bigg]
\end{eqnarray}
For the simplicity, we rewrite the $Q_4(h,\beta)$ as a polynomial function of $e^{\beta h}$ as
\begin{eqnarray}
\label{eqn:approx_lambda1}
&&  Q_{4}(h,\beta) = a_0e^{2\beta h}+b_0e^{\beta h}+d_0
\end{eqnarray}
Here, the system parameters $a_0, b_0, d_0$ are defined as
\begin{eqnarray}
\label{eqn:a0b0d0}
& a_0=e^{-\frac{\beta}{4}(J_q^{z}+J_c+4J_{cq}+4J_{d})}, \nonumber \\ 
& b_0=2e^{\frac{\beta}{4}(J_q^{z}-J_c)}Cosh[\frac{\beta J_q^{xy}}{2}]+4e^{-\frac{\beta}{4}(J_q^{z}-J_c)}, \nonumber \\
& d_0=2e^{\frac{\beta (J_q^{z}+J_c)}{4}}\bigg[ Cosh[\frac{\beta J_q^{xy}}{2}]+Cosh[\frac{\beta J_q^{xy}\Delta_2}{2}]\bigg] \nonumber \\
& +e^{-\frac{\beta}{4} (J_q^{z}+J_c-4J_{cq}-4J_{d})}.
\end{eqnarray}

\section{Results}
\label{sec:results}
We study the magnetization properties in the presence of a longitudinal ($h$) and transverse ($h^x$) magnetic field in Sec.\ref{sec:longitudinal} and Sec.\ref{sec:magx} respectively. In both of these cases, the studies are done for four sets of exchange parameters: (i) $J_q=0.2, J_d=2.0$ for the SRFM, (ii) $J_q=2.0, J_d=0.4$ for the AAFM, (iii) $J_q=2.0, J_d=1.0$ for the Dimer, and (iv) $J_q=2.0, J_d=1.6$ for the SLFM phases. It is to be mentioned that in all these cases, we consider $J_c$ and $J_{cq}$ to be unity. 
\subsection{Magnetization process in the presence of a longitudinal magnetic field}
\label{sec:longitudinal} 
\begin{figure}[t]
\centering
\includegraphics[width=1.0\linewidth]{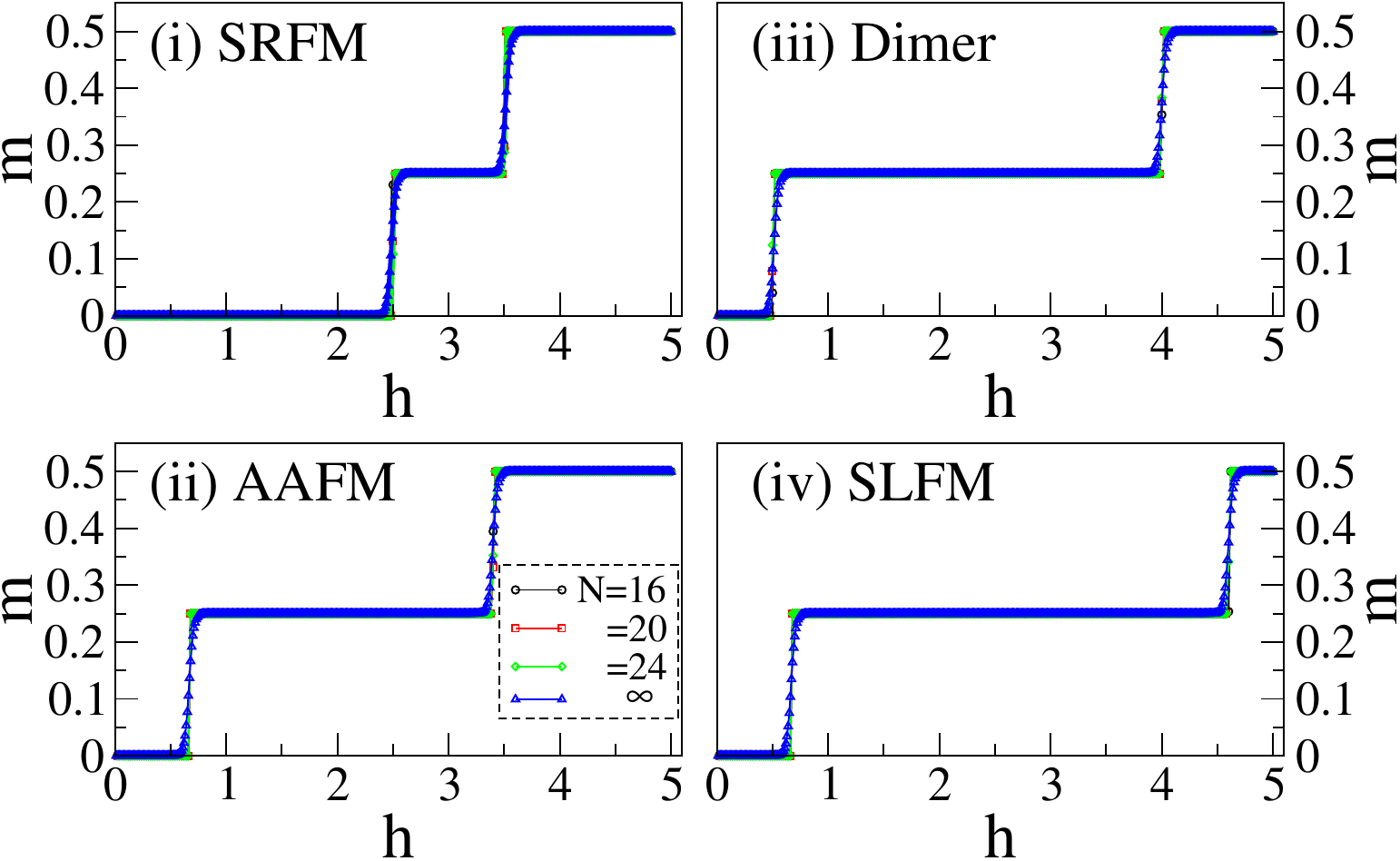}
\caption{(PBC) Black, red, and green colors are the magnetization per site in the presence of a longitudinal field $ h$ for the system sizes $N=16, 20, 24$ using ED. The blue color represents the magnetization calculated using the TM method at $T/J_{c}\rightarrow 0$. The values of $J_q$, $J_d$ are (i) $0.2, 2.0$ for the SRFM,  (ii) $2.0, 0.4$  for the AAFM,  (iii) $2.0, 1.0$  for the Dimer, and (iv) $2.0, 1.6$ for the SLFM phases respectively.}
\label{fig:magzpbc}
\end{figure}
\subsubsection{Magnetization vs field}
\label{sec:magz} 
We obtain the per-site magnetization $m$ for a finite system size $N$ by calculating the spin gap i.e., the difference between the low-lying states of two different spin $S^z$ sectors using ED.
In Fig.\ref{fig:magzpbc}.[(i)-(iv)], we show the finite size scaling of the m-h curve for three system sizes $N=16, 20, 24$ using ED. $m-h$ curve shows\ three plateau phases: $m=0, 1/4$, and $1/2$  connected by two magnetic jumps in each of the subfigures. The first jump is at $h_{c1}$ from $m=0$ to $1/4$, and the other at $h_{c2}$ from $m=1/4$ to full saturation of magnetization ($m=1/2$) in all four quantum phases as shown in Fig.\ref{fig:magzpbc}[(i)-(iv)]. The $h_{c1}$ takes values 2.5, 0.7, 0.5, and 0.7, whereas, $h_{c2}$ are 3.5, 3.5, 4, 4.5 for (i) the SRFM, (ii) the AAFM, (iii) the Dimer, and (iv) the SLFM phases respectively. Later on, we discuss in detail that these magnetic transitions show plateaus due to the spin gap and these jumps correspond to the overlap of the plateaus by unbinding of the rung dimers of equal energy. In all of the magnetization curves, it is noticed that there is a negligible finite-size effect. This is also noticed in the thermodynamic limit ($N \rightarrow \infty$) for the zero-temperature case ($\beta \rightarrow \infty$) using TM method.

Using the TM method, the per-site magnetization is obtained as $m=-\frac{1}{4}\frac{\partial F(h, \beta)}{ \partial h}$, where $F(h,\beta)$ is the free energy 
for one unit and can be defined as $F(h,\beta)=-\frac{1}{\beta} ln[Q_4(h,\beta)]$.  The per-site magnetization in general can be written as 
\begin{eqnarray}
\label{eq:mag_tm}
m=\frac{1}{4} \times \frac{[2a_0 e^{2 \beta h}+ b_0 e^{\beta h}]}{[a_0 e^{2 \beta h}+ b_0 e^{\beta h}+ d_0]}
\end{eqnarray}
From the above equation, it is clear that the exponential terms compete and dominate each other in different limits of fields and $m$ can take the discrete values $0, \frac{1}{4}, \frac{1}{2}$ (for $\beta \rightarrow \infty$). These magnetization plateaus can be connected by a few jumps which are associated with the critical fields and these critical fields are the mid-points of the jumps. At two critical fields $h_{c1}$ and  $h_{c2}$, $m$ takes the values $\frac{1}{8}$ and $\frac{3}{8}$ respectively. However, the critical fields can be obtained from the Eq.\ref{eq:mag_tm} as
\begin{eqnarray}
\label{eq:hc1hc2_simple}
h_{c1} \approx \frac{1}{\beta} ln\left[\frac{d_0}{b_0}\right],
h_{c2} \approx \frac{1}{\beta} ln\left[\frac{b_0}{a_0}\right].
\end{eqnarray} 
These critical fields are found to match with the exact calculations discussed above and shown in the $m-h$ curve in Fig.\ref{fig:magzpbc}. Plateau width can be obtained as 
\begin{eqnarray}
\label{eqn:plateauwidth}
    d=|h_{c2}-h_{c1}|\approx \frac{1}{\beta} ln\left[\frac{b_0^2}{a_0d_0}\right]
\end{eqnarray}
From the above discussion one can conclude that in $\beta \rightarrow \infty$ limit, the partition function $Q_4(h,\beta)$ of Eq.\ref{eqn:approx_lambda1} can be simplified as following
\begin{equation}
\label{cases1}
Q_4(h,\beta) \approx \cases{d_0 & if $h \le h_{c1}$\\
b_0 e^{\beta h} & if $h_{c1} \le h \le h_{c2}$\\
a_0 e^{2\beta h} & if $h \ge h_{c2}$\\
}
\end{equation}
Similarly, the free energy $F(h,\beta)$ can be written for three different plateau phases
\begin{equation}
F(h,\beta) =
\cases{
        -\frac{1}{\beta}ln\big[d_0\big] & if $h \le h_{c1}$\\
        -\frac{1}{\beta}ln\big[b_0\big]-h & if  $h_{c1} \le h \le h_{c2}$\\
        -\frac{1}{\beta}ln\big[a_0\big]-2h & if $h \le h_{c1}$\\
}
\end{equation}
From the above expressions of the free energy, it is noticed that the per-site magnetization can easily be obtained from $m=-\frac{1}{4}\frac{\partial F(h,\beta)}{\partial h}$ to be $0,1/4, 1/2$ for the three plateau phases.  All of the four nonmagnetic GS phases (i.e., $m=0$) have a finite magnetic excitation gap and it requires a finite external magnetic field $h$ to reach a magnetic GS.  The lowest magnetic excitation is in $m=1/4$, for which the gap from the nonmagnetic GS is closed by a finite field $h_{c1}$. However, the $m=1/4$ state has also a finite magnetic excitation gap and it requires a field $h_{c2}$ to achieve the higher magnetic excitation $m=1/2$, which is a fully polarized phase (FP). All the nonmagnetic GS ($m=0$), $m=1/4$ magnetic GS and $m=1/2$ magnetic GS phases can be described by the system parameters $d_0$, $b_0$, and $a_0$ respectively. 
\begin{figure}[t]
\centering
\includegraphics[width=1.0\linewidth]{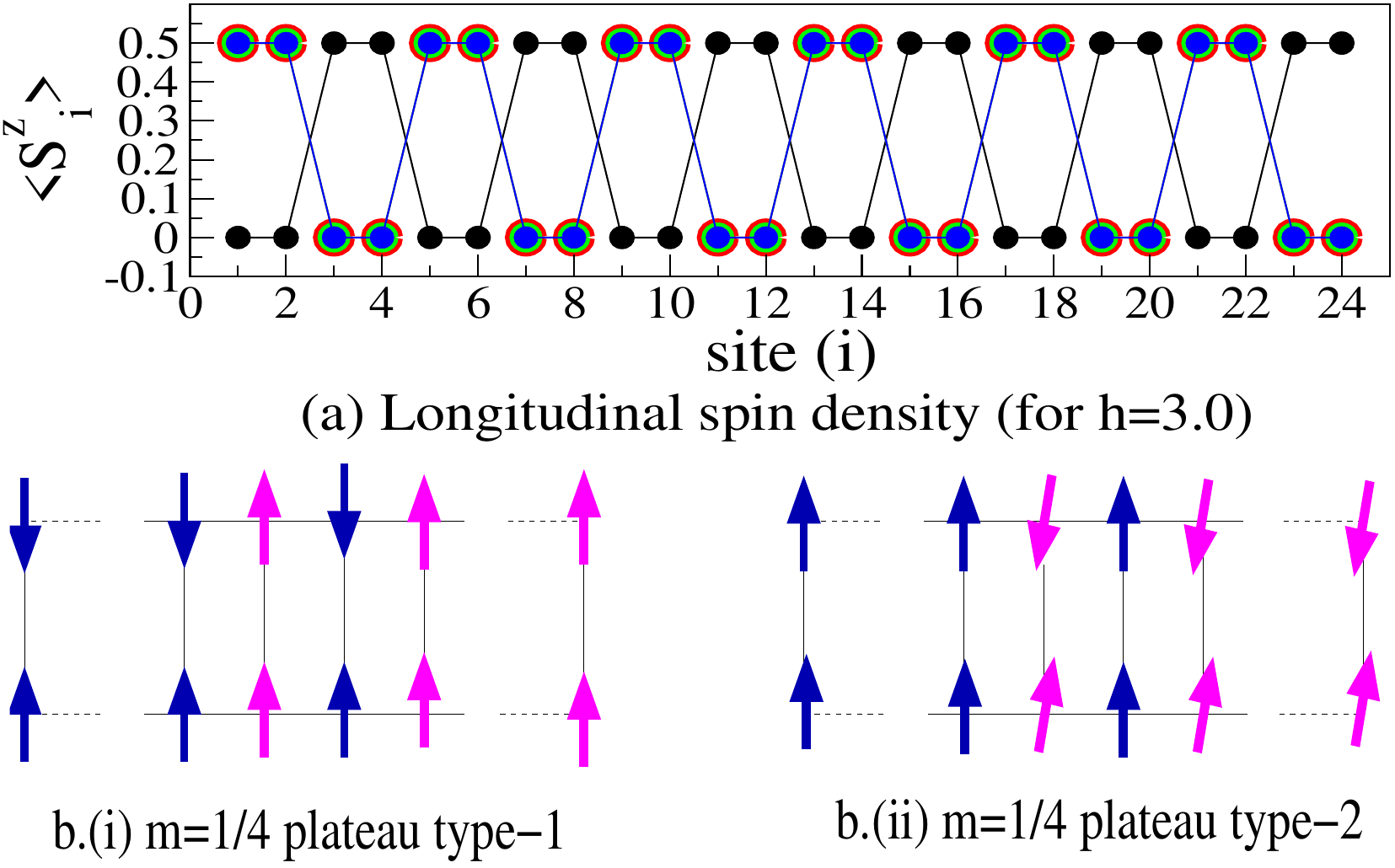}
\caption{(a) Longitudinal spin density $S^z_i$ is shown as a function of site index $i$ for four GS phases: the SRFM ($J_q=0.2$,$J_d=2.0$), the AAFM ($J_q=2.0$,$J_d=0.4$), the Dimer ($J_q=2.0$,$J_d=1.0$), and the SLFM ($J_q=2.0$,$J_d=1.6$) using black, red, green, blue colors respectively. The spin configurations in b.(i) $m=1/4$ plateau type-1, b.(ii) $m=1/4$ plateau type-2 are shown. In  $m=1/4$ plateau type-1, all the Heisenberg rung pairs $S-S$ (magenta spins) are fully polarized, and in $m=1/4$ plateau type-2, all the Ising rung pairs $\sigma-\sigma$ (blue spins) are fully polarized.  }
\label{fig:configuration}
\end{figure}

To investigate the nature of all the plateau phases, spin density $\langle S^z_i\rangle$ at site $i$ calculated using the ED is shown in Fig.\ref{fig:configuration}. (a). The SRFM phase is stabilized in the presence of the dominant diagonal Ising exchange whereas, in the other three phases isotropic Heisenberg rung exchange is dominant. Therefore, the formation of the $m=1/4$ plateau from the SRFM phase (with low $J_q$ and high $J_d$) is different from the other three phases. In the SRFM phase, the diagonal exchange $J_d$ exhibits Ising characteristics and is stronger compared to the other exchanges. The formation of rung dimers and the dominant Ising exchange create a finite gap between the non-magnetic ($m=0$) and magnetic ($m=1/4$) phases. To close this gap and transition to the $m=1/4$ plateau phase, a large field $h_{c1}=2.5$ is required, as shown in Fig. \ref{fig:magzpbc}. (i). At this critical field, the weakly coupled Heisenberg rung dimers ($S-S$) break down and get polarized, resulting in a finite spin density of 0.5, as depicted in Fig. \ref{fig:configuration}.(a). The ground state spin configuration for the $m=1/4$ plateau is termed ``m=1/4 plateau type-1" or ``P-I" and is shown in Fig. \ref{fig:configuration}.b.(i). With further increase of the field around $h=3.5$, all the spin pairs are broken and the system goes to the FP phase. 

The AAFM, Dimer, and SLFM phases are characterized by strong Heisenberg rung exchange $J_q$, leading to the formation of strong dimers on $S-S$ pairs, which are energetically more stable compared to the dimers of $\sigma-\sigma$ pairs interacting with Ising exchange. In the AAFM phase, with anisotropic antiferromagnetic spin alignment on the ladder for $J_q=2.0$ and $J_d=0.4$, a very small field $h_{c1}=0.7$ is sufficient to break the $\sigma-\sigma$ pairs and reach the $m=1/4$ plateau, as shown in Fig.\ref{fig:magzpbc}.(ii). In the $m=1/4$ plateau of AAFM, the rung spins of $\sigma-\sigma$ pairs are fully polarized with a spin density of 0.5, as depicted in Fig.\ref{fig:configuration}.(a). The spin configuration of this type of magnetic phase is referred to as ``m=1/4 plateau type-2" or ``P-II" and is shown in Fig. \ref{fig:configuration}.b.(ii).

In the Dimer phase, all the rung dimers are isolated, and the Ising dimers have a smaller spin gap as compared to the $S-S$ rung singlets. An external field of $h_{c1}=0.5$ is sufficient to close the gap for a given value of $J_q=2.0$ and  $J_c=1.0$ as shown in Fig.\ref{fig:magzpbc}.(iii). Due to the formation of perfect dimers through strong $J_q$ exchange in this phase, the P-II phase exhibits a much larger gap, leading to a broader plateau width compared to the AAFM phase, as shown in Fig. \ref{fig:magzpbc}.(iii). In the SLFM phase, the ground state exhibits ferromagnetic spin arrangements along the leg, while spins on different legs are oppositely aligned. Fig. \ref{fig:magzpbc}.(iv) demonstrates that the onset of the P-II occurs at a field of $h_{c1}=0.7$, with the largest plateau width observed in this phase. The spin density analysis reveals that the dimers formed by $\sigma-\sigma$ pairs become polarized along the field direction in the P-II phase as shown in Fig.\ref{fig:configuration}.b.(ii). Similar to the AAFM phase, the $\sigma-\sigma$ rung dimers are weaker in this phase also, while the oppositely aligned legs enhance the stability of singlets on the Heisenberg pair $S-S$, resulting in an extended plateau width. With further increase in the applied field, the P-II plateau as shown in Fig.\ref{fig:configuration}.b.(ii) is disrupted, and a sharp jump occurs at $h_{c2}=4.5$ and reaches the FP plateau. In all four phases, either all $S-S$ or all $\sigma-\sigma$ rung pairs are simultaneously disrupted, leading to magnetic jumps.
\subsubsection{Quantum Fidelity and Bipartite Concurrence}
\label{sec:quant_fid}
The spin arrangement of the plateau phases is quite different for all four phases at $h=0$. Therefore, we expect the wave function of the plateau phases should be different and can be understood from the perspective of quantum information theory. 
In this subsection, we calculate and show the zero temperature limit quantum fidelity and bipartite concurrence to analyze the plateaus, and these can be obtained from the partition function as discussed below.  Quantum fidelity is the measurement of overlap between two states and can be used to characterize the phase transition on the tuning of parameters. Quan et. al. have shown that fidelity can be obtained from the partition function \cite{PhysRevE.79.031101}. Similarly, we calculate the quantum fidelity for the field $h$ with small perturbation $\delta h$ as: 
\begin{eqnarray}
\label{eq:fidel_partition}
\mathcal{F}(h,\beta)=\frac{Q_4(h,\beta)}{\sqrt{Q_4(h+ \delta h, \beta)Q_4(h-\delta h, \beta)}}
\end{eqnarray} 
 where $Q_4(h,\beta)$ is the partition function for one unit cell in this case. $\mathcal{F}(h, \beta)$ is unity when there is a unique state and discontinuous at the phase transition points. The Field fidelity susceptibility $\mathcal{\chi_F}(h, \beta)=\frac{\partial \mathcal{F}(h, \beta)}{\partial h}$ is zero in unique state and it diverges at the transition.
\begin{figure}[t]
\centering
\includegraphics[width=1.0\linewidth]{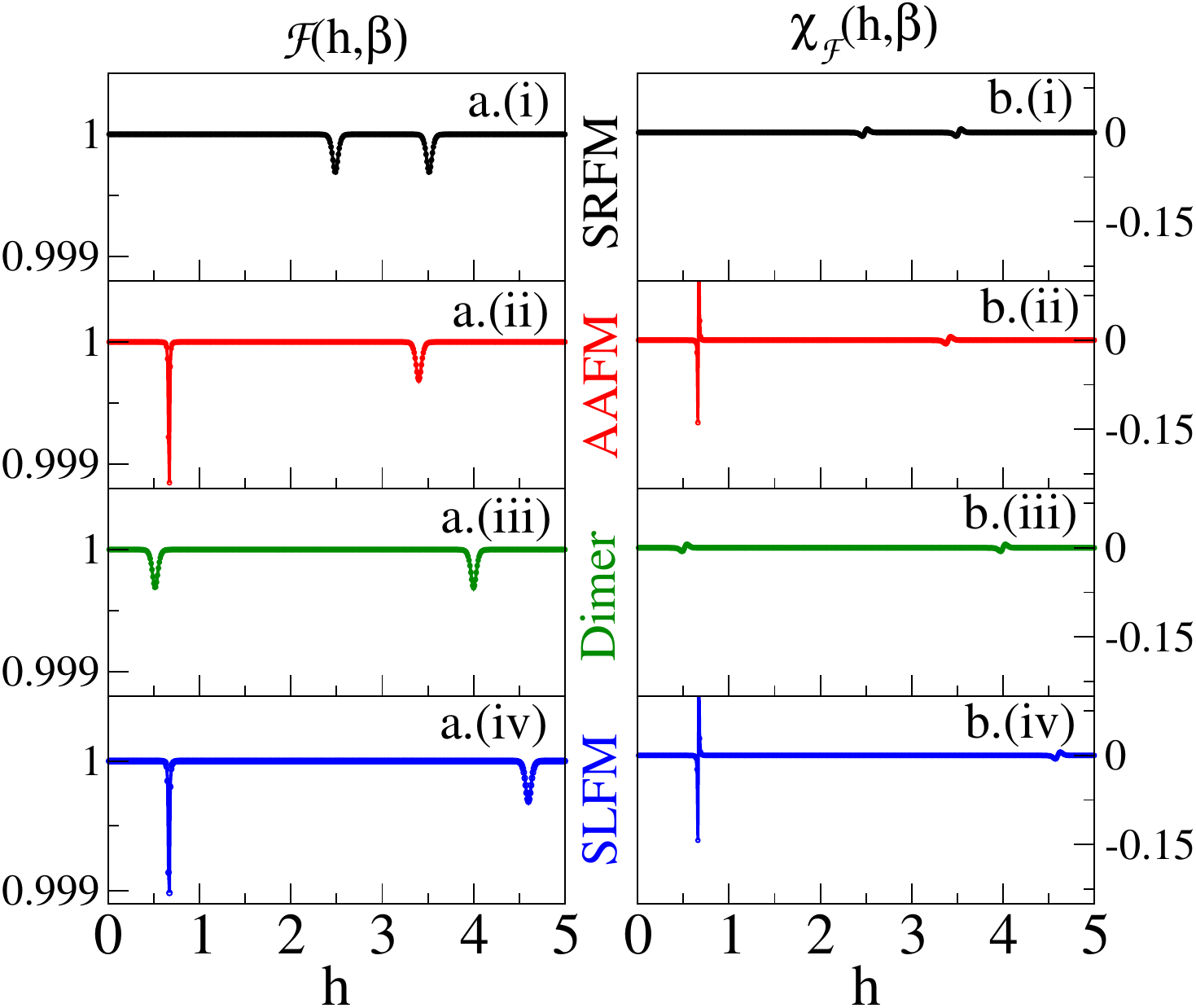}
\caption{ (a) (left column) Quantum fidelity $\mathcal{F}(h,\beta)$ and (b) (right column) fidelity susceptibility   $\mathcal{\chi_F}(h, \beta)$ calculated using TM are shown for the thermodynamic limit ($N \rightarrow \infty$) at $T/J_c\rightarrow 0$. Black, red, green, and blue colors representing four phases: the SRFM ($J_q=0.2$,$J_d=2.0$), the AAFM ($J_q=2.0$,$J_d=0.4$), the Dimer ($J_q=2.0$,$J_d=1.0$), and the SLFM ($J_q=2.0$,$J_d=1.6$) respectively are arranged sequentially from top to bottom in both the columns.}
\label{fig:fflvsh}
\end{figure}
$\mathcal{F}(h,\beta)$ can further be written in terms of the free energy $F(h,\beta)$ and magnetic susceptibility  $\chi(h,\beta)$ as 
\begin{eqnarray}
\label{eq:fidel_chi}
& \mathcal{F}(h,\beta)=e^{-\beta \big[ F(h,\beta) -\frac{F(h -\delta h,\beta)+F(h +\delta h,\beta)}{2} \big]} \nonumber \\
& \approx e^{-\big[\frac{(\beta \delta h)^2}{4} \chi(h,\beta)\big]}
\end{eqnarray}
In the plateau phases, $\chi(h,\beta)=-\frac{\partial^2{F(h,\beta)}} {\partial{h^2}}=\frac{\partial m}{\partial h}$ is zero and so the fidelity $\mathcal{F}(h,\beta)$ in Eq.\ref{eq:fidel_chi} is unity. On the other hand, at the magnetic jumps, the $\chi(h,\beta)$ diverges, and the $\mathcal{F}(h,\beta)$ decreases from unity. 

Fig.\ref{fig:fflvsh}.a.[(i)-(iv)], and b.[(i)-(iv)] show the plot of $\mathcal{F}(h, \beta)$ (left column) and $\mathcal{\chi_F}(h, \beta)$ (right column) as a function of $h$ respectively for four different phases: (i) the SRFM, (ii) the AAFM, (iii) the Dimer, and (iv) the SLFM respectively. In each of the sub-figures of
$\mathcal{F}(h, \beta)$  and $\mathcal{\chi}(h, \beta)$, two discontinuities are noticed for all four phases. All of these discontinuities are consistent with the jumps of the $m-h$ curve in Fig.\ref{fig:magzpbc} and represent the magnetic phase transitions. 

We also calculate the bond order to understand the configurational change and bipartite concurrence to measure the quantum nature of the $S-S$ pair which is connected through a Heisenberg rung exchange $J_q$. If the concurrence has some finite value, the wavefunction is in a mixed or entangled state, otherwise, it is in a pure state if the concurrence is zero.  Wooters et.al. and Karlova et. al. in their study calculate concurrence for a spin pair connected by Heisenberg interaction in terms of the local pair magnetization and spatial correlations: longitudinal, transverse to detect phase transitions at a finite temperature \cite{PhysRevLett.80.2245, karlova2020}. In our study, we calculate the bipartite concurrence $\mathcal{C}(J_q,h)$ for the Heisenberg rung pair connected by exchange interaction $J_q$=($J_q^z$, $J_q^{xy}$) in presence of the longitudinal field as
\begin{eqnarray}
\label{eq:bipartite}
&\mathcal{C}(J_q,h)=max \Big\{0,
4 |{c^T(J_q,h)}|- \nonumber \\
&2\sqrt{[\frac{1}{4}+c^L(J_q,h)]^2-
\big[m'(J_q,h)\big]^2}
\Big\}
\end{eqnarray} 
Where, $m'(J_q,h)$, $c^L(J_q,h)$ and $c^T(J_q,h)$ are the pair magnetization, the longitudinal and transverse component of bond order respectively, and these can be defined as:\\
\begin{eqnarray}
\label{eq:pairjq}
&  m'(J_q,h)=\frac{1}{2}\big<S^z_{2j,1}+S^z_{2j,2}\big>=-\frac{1}{2}\frac{\partial{F_1(h,h',\beta)}}{\partial{h^{'}},}  \\
& c^L(J_q,h)=\big<S^z_{2j,1}S^z_{2j,2}\big>=-\frac{1}{\beta}\frac{\partial{[\ln{Q_4(h,\beta)}]}}{\partial{J_q^{z}}}, \\
& c^T(J_q,h)=\big<S^x_{2j,1}S^x_{2j,2}\big>=-\frac{1}{2\beta}\frac{\partial{[\ln{Q_4(h,\beta)}]}}{\partial{J_q^{xy}}.}
\end{eqnarray}
Where, $F_1(h,h',\beta)$ is the free energy of one unit for which $h, h'$ are the applied magnetic fields on the $\sigma, S$ spins respectively. For the case,  $h^{'}=h$, $F(h,\beta)=F_1(h, h',\beta)$. For a better understanding of the spin arrangement of the $\sigma$ spins within the unit, we calculate the longitudinal bond order $c^L(J_c,h)$ for the $\sigma-\sigma$ and $c^L(J_d,h)$ for the diagonal $\sigma-S$ spin pairs as:
\begin{eqnarray}
\label{eq:pairjc}
&  c^L(J_c,h)=\big<\sigma_{2j+1,1}\sigma_{2j+1,2}\big>=-\frac{1}{\beta}\frac{\partial{[\ln{Q_4(h,\beta)}]}}{\partial{J_c}}  \\
&c^L(J_d,h)=\big<\sigma_{2j+1,1} S^z_{2j+1,2}\big>=-\frac{1}{4 \beta}\frac{\partial{[\ln{Q_4(h,\beta)}]}}{\partial{J_d}}
\end{eqnarray}
\begin{figure}
\centering
\includegraphics[width=1.0\linewidth]{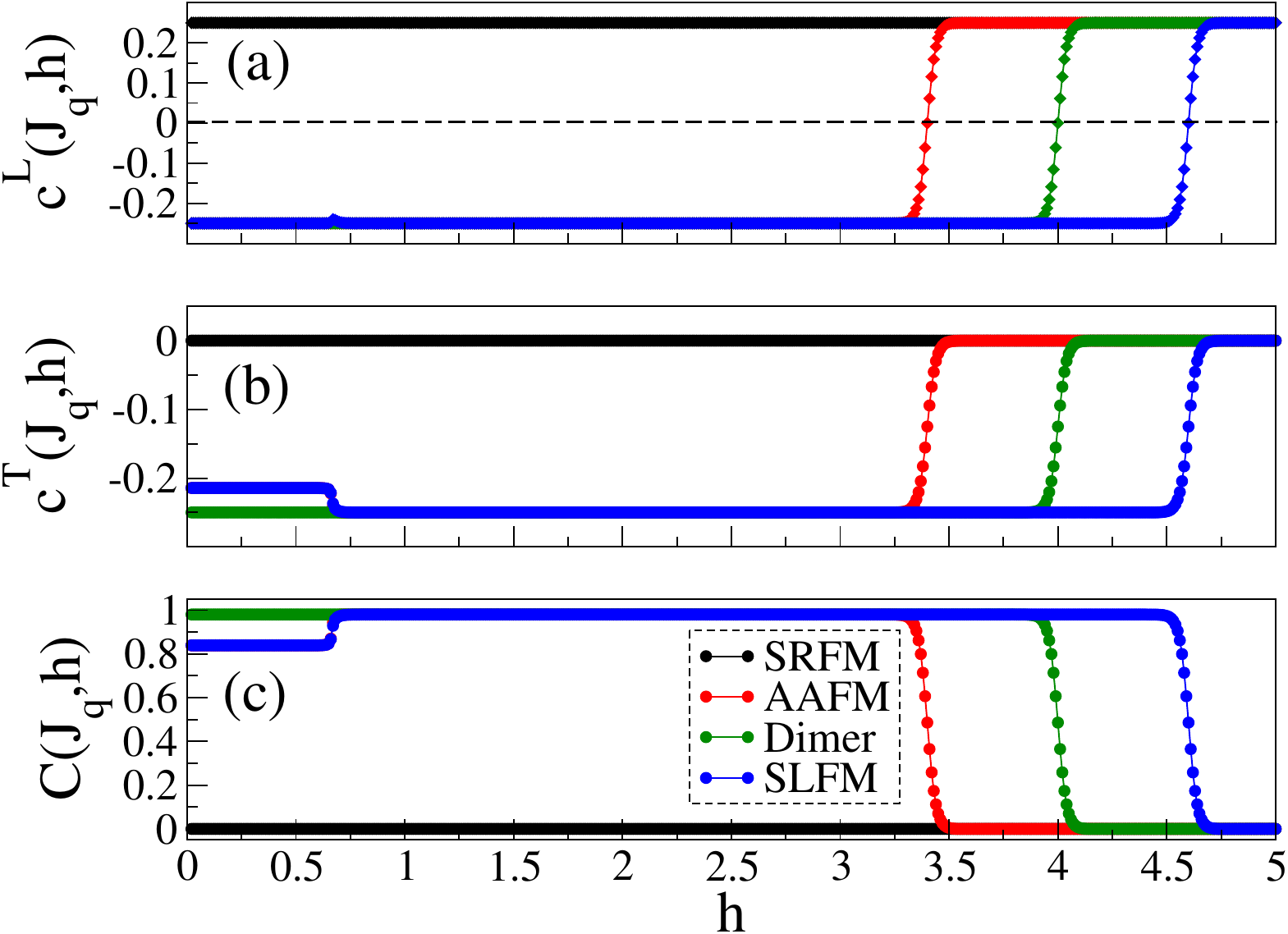}
\caption{(a) Longitudinal bond order $c^L(J_q,h)$, (b) transverse bond order $c^T(J_q,h)$, (c) Quantum concurrence $\mathcal{C}(J_q,h)$ for the Heisenberg spin pair $S-S$ connected by rung strength $J_q$ are shown as a function $h$. Black, red, green, and blue colors represent four phases: the SRFM ($J_q=0.2$,$J_d=2.0$), the AAFM ($J_q=2.0$,$J_d=0.4$), the Dimer ($J_q=2.0$,$J_d=1.0$), and the SLFM ($J_q=2.0$,$J_d=1.6$) respectively at $T/{J_c}=0.02$.}
\label{fig:concvsh}
\end{figure}
However, the  $c^L(J_q,h)$, $c^T(J_q,h)$, and the $\mathcal{C}(J_q,h)$ are shown in Fig.\ref{fig:concvsh}. (a), (b) and (c) respectively for all four phases (i) the SRFM, (ii) the AAFM, (iii) the Dimer, and (iv) the SLFM. Fig.\ref{fig:concvsh} can be analyzed separately for different GS phases by approximating the partition function $Q_4(h,\beta)$ of Eq.\ref{eqn:approx_lambda1}, as we mentioned earlier in Sec.\ref{sec:magz} that $d_0, b_0$, and $a_0$ can describe the $m=0, 1/4$, and $1/2$ plateau phases respectively for different limits of exchange parameters $J_q, J_d$, and field $h$. 

The partition function for the SRFM phase having lower $J_q$ and higher $J_d$ can be written for three plateau phases: $m=0, 1/4, 1/2$ as below-
\begin{eqnarray}
Q_4(h,\beta) \approx
\cases{
e^{-\frac{\beta}{4} (J_q^{z}+J_c-4J_{cq}-4J_{d})} &  m=0\\
       4e^{-\frac{\beta}{4} (J_q^{z}-J_{c}-4h)} &   m=1/4\\
      e^{-\frac{\beta}{4} (J_q^{z}+J_{c}+4J_{cq}+4J_{d}-8h)} &  m=1/2
}
\end{eqnarray}
In all of the plateaus, $c^L(J_q,h)=\frac{1}{4}$ and $c^T(J_q,h)=0$ emphasizes the Ising dominance with the ferromagnetic alignment of the spins along the Heisenberg rung as shown in Fig.\ref{fig:concvsh}.(a), and (b) respectively. The ferromagnetic alignment suggests $m^{'}(J_q,h)=\pm 1/2$ and so the Eq.\ref{eq:bipartite} demands the $\mathcal{C}(J_q,h)$ to be always zero, which in other words can be thought of as there is no quantum concurrence or no quantum entanglement between the two $S$ spins as shown in Fig.\ref{fig:concvsh}. (c). It is to be noticed that $m^{'}(J_q,h)=\pm 1/2$ represents the P-I phase as shown in Fig.\ref{fig:configuration}.b.(i). It can now be stated that the P-I phase is a pure state.

For the other three phases: the AAFM, the Dimer, and the SLFM with sufficiently larger $J_q$, the magnetization can be analyzed from the $Q_4(h,\beta)$ for various limits of the system parameter $\Delta_2$. We discuss the quantum concurrence for the plateau phases by writing down the partition functions separately.

\underline{(i) $m=0$ plateau:} \\
The partition function in this phase can be written as
\begin{eqnarray}
\label{eq:zerodelta}
&Q_4(h,\beta) \approx \nonumber \\
&2e^{\frac{\beta (J_q^{z}+J_c)}{4}}\bigg[ Cosh[\frac{\beta J_q^{xy}}{2}]+Cosh[\frac{\beta J_q^{xy}\Delta_2}{2}]\bigg]
\end{eqnarray}
The longitudinal bond order $c^L(J_q,h)$ in this case is  $-1/4$ as shown in Fig.\ref{fig:concvsh}.(a) whereas, the transverse bond order $c^T(J_q,h)$ is controlled by the interplay of the two terms in the partition function. For the Dimer phase ($\Delta_2=1$), the $c^T(J_q,h)=-1/4$, whereas, for the AAFM and SLFM phases ( $\Delta_2 > 1$), $c^T(J_q,h)$ is less in magnitude as compared to the Dimer phase as shown in Fig.\ref{fig:concvsh}. (b). The large transverse correlation along the Heisenberg spin pairs results in large $\mathcal{C}(J_q,h)$ close to unity for the AAFM and SLFM phases and it is maximum (unity) for the Dimer phase as shown in Fig.\ref{fig:concvsh}.(c).

\underline{(ii) $m=1/4$ plateau}\\
The approximated partition function in this magnetic phase is 
\begin{eqnarray}
\label{eq:halfdelta}
&Q_4(h,\beta) \approx 2e^{\frac{\beta}{4}(J_q^{z}-J_c+4h)}Cosh[\frac{\beta J_q^{xy}}{2}]
\end{eqnarray}
From the above partition function, one can obtain $c^L(J_c,h)=1/4$, which implies the polarization of the $\sigma$ spin pairs along the longitudinal field and it represents the P-II phase. In this plateau phase, the Heisenberg spin pairs have the longitudinal  $c^L(J_q,h)$ and the transverse bond order $c^T(J_q,h)$ both as $-1/4$ and it is shown in Fig.\ref{fig:concvsh}.(a), and (b). The maximum value of  $c^T(J_q,h)$ in this phase results in giving the maximum concurrence  $\mathcal{C}(J_q,h)$ equal to unity for all the phases: (i) the AAFM, (ii) the Dimer, and (iii) the SLFM as shown in Fig.\ref{fig:concvsh}. (c). The Eq.\ref{eq:halfdelta} has no $J_d$ dependence implying the Ising and HB rungs are isolated. From these results, one can remark that the P-II is maximally entangled due to the strong singlet formation along the HB rung.
\\
 \underline{(iii) $m=1/2$ plateau}\\
 In this plateau phase, the partition function can be written as 
 \begin{eqnarray}
\label{eq:fpdelta}
&Q_4(h,\beta) \approx e^{-\frac{\beta}{4}(J_q^{z}+J_c+4J_{cq}+4J_{d}-8h)},
\end{eqnarray}
which is exactly the same as discussed before for the SRFM phase. The exponent of the partition function suggests that all the longitudinal bond orders $c^L(J_q,h), c^L(J_c,h)$, and $c^L(J_d,h)$ are $1/4$ and it is the signature of the FP phase with vanishing transverse bond order along the Heisenberg exchange $J_q$. The maximum longitudinal bond order and vanishing transverse bond order in the FP phase ensure that it is a pure state.

\subsubsection{Quantum phase diagrams}
\label{sec:magnetic_phase_diagrams}
\begin{figure}
\includegraphics[width=1.0\linewidth]{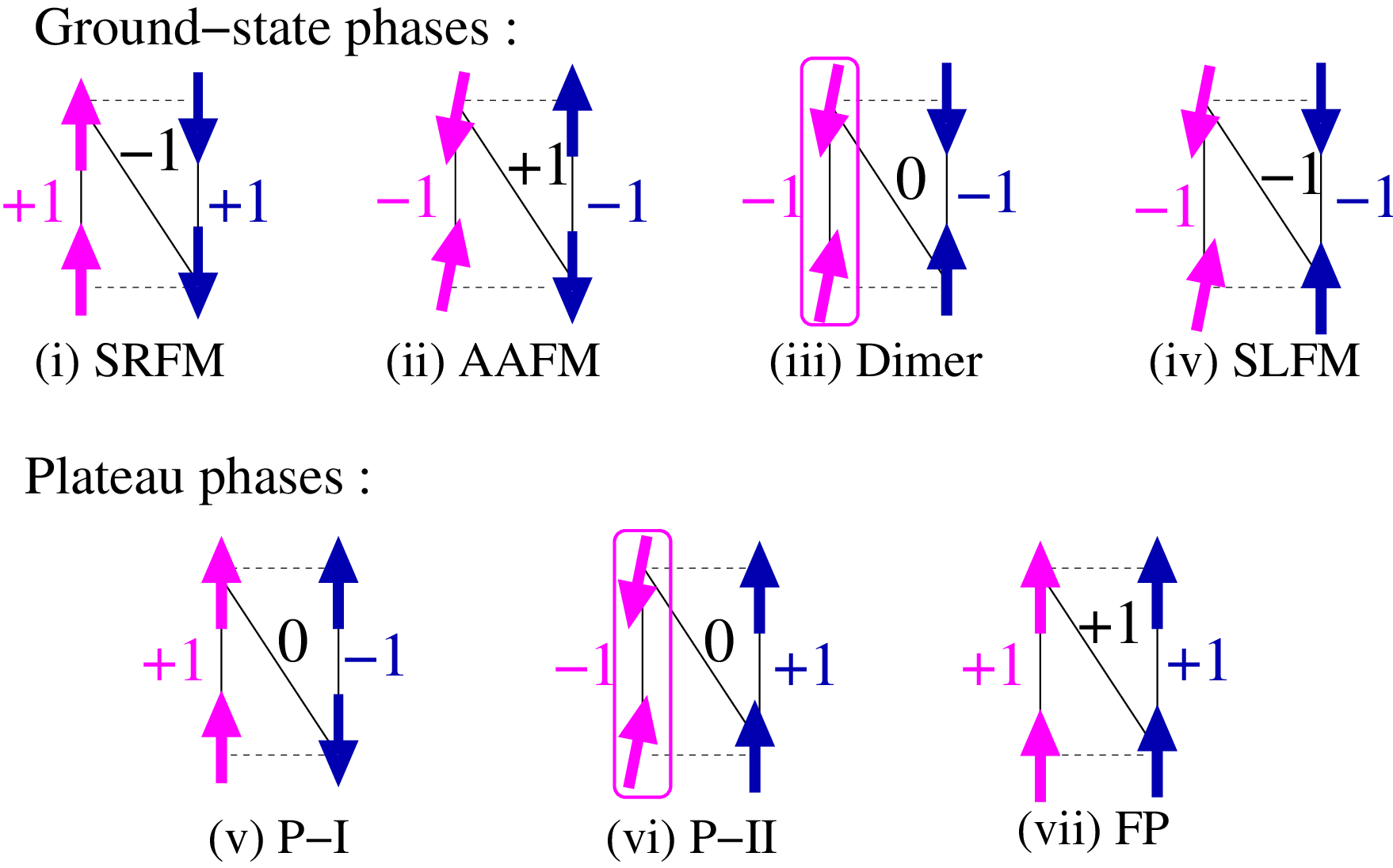}
\caption{Schematics of the GS and the plateau phases are shown. Sign of the bond orders with exchange interactions $J_q$,$J_d$, and $J_c$ are shown in magenta, black, and blue colors respectively. Boxes represent the perfect singlet Dimer formation.}
\label{fig:QPI}
\end{figure}

In this section, we analyze the quantum phases as a function of $J_d$ and $J_q$ at different longitudinal magnetic fields $h=0, 1, 2$, and $3$. For the sake of completeness, we show the $h=0$ quantum phases (i.e., the zero field GS phases), which are studied earlier in \cite{Saniur_Rahaman_2021}. To characterize the quantum phases uniquely, we define the quantum phase index (QPI) which is a function of three different bond orders within the unit cell of the lattice and can be written as
\begin{eqnarray}
\label{eqn:qpi}
&QPI=2Sign[c^L(J_q,h)]+Sign[c^L(J_d,h)]\nonumber \\
&+Sign[c^L(J_c,h)] 
\end{eqnarray}
Where $Sign[x]$ is the signum function and can be defined as
\begin{equation}
Sign[x]=
\cases{
        1 & for  $x > 0$\\
        0 & for  $x = 0$\\
        -1 & for $x < 0$\\
}
\end{equation}
The spin arrangements and the signum values of the longitudinal bond orders along Heisenberg rung $c^L(J_q,h)$, diagonal spin pair $c^L(J_d,h)$, and Ising rung $c^L(J_c,h)$ are shown for the quantum phases in Fig.\ref{fig:QPI}. 
\begin{table}[t]
\caption{\label{tab:table}
Quantum Phase Index (QPI) for different phases.}
\begin{indented}
\lineup
\item[]\begin{tabular}{@{}*{5}{l}}
\br                              
Phases & $Sign$ & $Sign$ & $Sign$ & $QPI$\\
 & $[c^L(J_q,h)]$ & $[c^L(J_d,h)]$ & $[c^L(J_c,h)]$\cr 
\mr
(i) SRFM & +1 &-1 &+1 & 2\cr
(ii) AAFM &-1 &+1 &-1 & -2\cr
(iii) Dimer &-1 &0 &-1 & -3\cr
(iv) SLFM &-1 &-1 &-1 & -4\cr
(v) P-I &+1 &0 &-1 & 1\cr
(vi) P-II &-1 &0 &+1 & -1\cr
(vii) FP &+1 &+1 &+1 & 4\cr
\br
\end{tabular}
\end{indented}
\end{table}
The signum value of the bond orders and the corresponding unique QPI for different quantum phases are tabulated in Table.\ref{tab:table}.

\begin{figure}[t]
\centering
\includegraphics[width=1.0\linewidth]{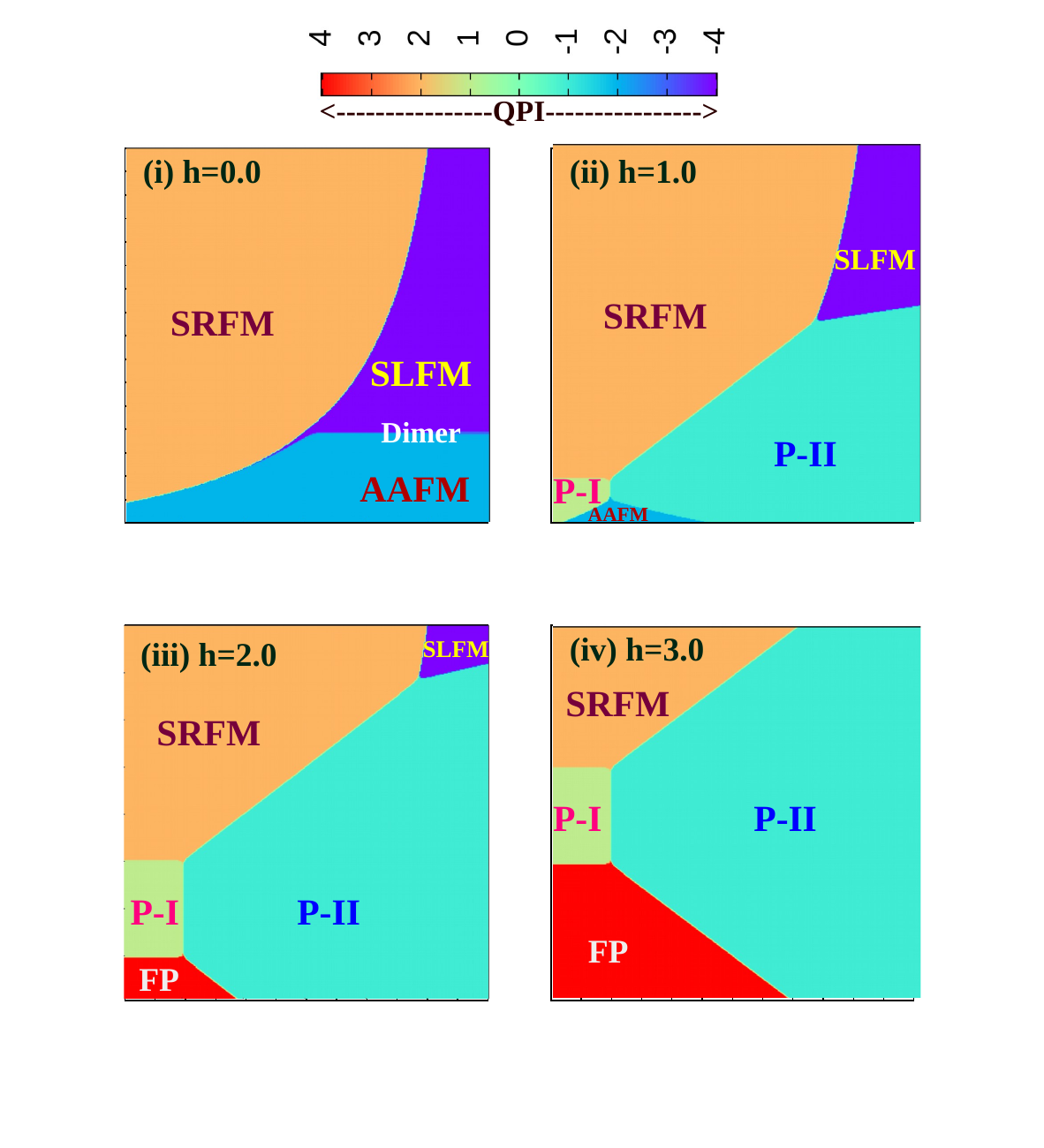}
\caption{Quantum Phase Index (QPI) is shown for five longitudinal fields $h=0, 1, 2, 3$ from (i)-(iv) in sequence. The color bar shows the QPI as defined in Eq.\ref{eqn:qpi}. The QPI values $2, -2, -3, -4, 1, -1, 4$ represent quantum phases: the SRFM, the AAFM, the Dimer, the SLFM, the P-I, the P-II, and the FP phases respectively.}
\label{fig:magnetic_phases}
\end{figure}

We analyze different plateau phases in the plane of $J_q-J_d$ for different values of $h$ as shown in Fig.\ref{fig:magnetic_phases}.(i)-(iv ). 
As it is seen in Fig.\ref{fig:magnetic_phases}.(i), $h=0$ phase diagram represents four GS phases: (i) the SRFM, (ii) the AAFM, (iii) the Dimer, and (iv )the SLFM with the four distinct QPI values as 2,-2,-3, and -4 respectively. Fig. \ref{fig:magnetic_phases}(ii) shows a few plateau phases along with the non-magnetic GS phases for the field $h=1.0$. In this phase diagram, the P-I and P-II phases are shown with QPI numbers 1, and -1 respectively. With the low $J_q$ value, the SRFM phase transit to P-I, whereas, in the vicinity of the SRFM-SLFM phase boundary, the regime with comparable $J_q$ and $J_d$ transit to P-II phase which is interesting and can be analyzed by writing down the critical fields for different limits of $J_q$. The spin gap or the critical field $h_{c1}$ for the SRFM phase in different $J_q$ limits can be obtained as
\begin{eqnarray}
\label{eqn:hc1srfm}
h_{c1} \approx
\cases{
         (J_d+1/2) & for $J_q<J_c$  \\
        (1+J_d-\frac{J_q^z}{2}-\frac{J_q^{xy}}{2}) & for $J_q>J_c$  \\
}
\end{eqnarray}
From the above equation, by considering $h_{c1}=1$, the phase boundaries between the SRFM-P-I and the SRFM-P-II phases can be obtained as $J_d=0.5$ and $J_d \approx J_q$ respectively and these boundaries are noticed in Fig.\ref{fig:magnetic_phases}. (ii).

Similarly, the spin gap $h_{c1}$ for the AAFM, the Dimer, and the SLFM phases can be written as a function of $J_q$ and $\Delta$ in the following manner
\begin{eqnarray}
\label{eqn:hc1aafm}
h_{c1} \approx
         \frac{1}{2}+\frac{1}{\beta}ln\big[1+\frac{Cosh[\frac{\beta J_q^{xy} \Delta_2}{2}]}{Cosh[\frac{\beta J_q^{xy}}{2}]} \big]
\end{eqnarray}
The critical fields $h_{c1}$ in Fig.\ref{fig:magzpbc}. (ii)-(iv) for the AAFM (0.7), the Dimer (0.5), and the SLFM (0.7) phases agree well with the above equation. Also, from the Eq.\ref{eqn:hc1aafm}, one can determine the phase boundaries between the AAFM-P-II, and the SLFM-P-II as a function of system parameters  $J_q$, $J_d$ or $\Delta_2$. However, the major part of the parameter space in Fig.\ref{fig:magnetic_phases}. (iii) is occupied by P-II due to the low spin gap in the AAFM, the Dimer, and the SLFM phases. In Fig. \ref{fig:magnetic_phases}(iii), the P-II occupies the largest area and replaces the AAFM phase completely for the field $h=2$. A small area of the FP phase emerges in this case for very small values of $J_q$ and $J_d$ in Fig.\ref{fig:magnetic_phases}.(iii).

The FP phase appears either from the P-I or P-II phase on the application of the critical field $h_{c2}$. This critical field can be written as a function of the system parameters in the following way 
\begin{eqnarray}
\label{eqn:hc2srfm}
h_{c2} \approx
    \cases{
         (J_d+3/2) & for P-I phase  \\
        \frac{(J_q^z+2J_d+2)}{2}  & for P-II phase  \\
    }
\end{eqnarray}
From the above equation, by taking $h_{c2}=2$, one can determine the phase boundary between the P-I-FP phase as $J_d=1/2$. Similarly, the P-II-FP phase boundary can also be obtained from the above equation. For $h=3$ in \ref{fig:magnetic_phases}. (iv), the SLFM phase is replaced by P-II, whereas, for the large $J_d$ value, the SRFM phase still exists in the magnetic phase diagram due to the largest spin gap $h_{c2}=(J_d+3/2)$ as in Eq.\ref{eqn:hc2srfm}. With an even larger field value, all other phases vanish and the FP phase occupies the phase diagram completely.
\subsection{Magnetization process in the presence of a transverse field}
\label{sec:magx}
Using many experimental techniques in general, this kind of system is synthesized in powder form or single crystal form and so to understand the directional dependence of the field on the magnetization, we study the effect of the transverse field $h^x$ in our model \cite{guo2020crystal,kakarla2021single}. In the presence of $h^x$, the Hamiltonian of different units do not commute to each other, and therefore, the TM method never works, we use the ED method to show the finite size scaling of magnetization for three system sizes and then analyze the spin density for $N=24$. 
\begin{figure}[t]
\centering
\includegraphics[width=1.0\linewidth]{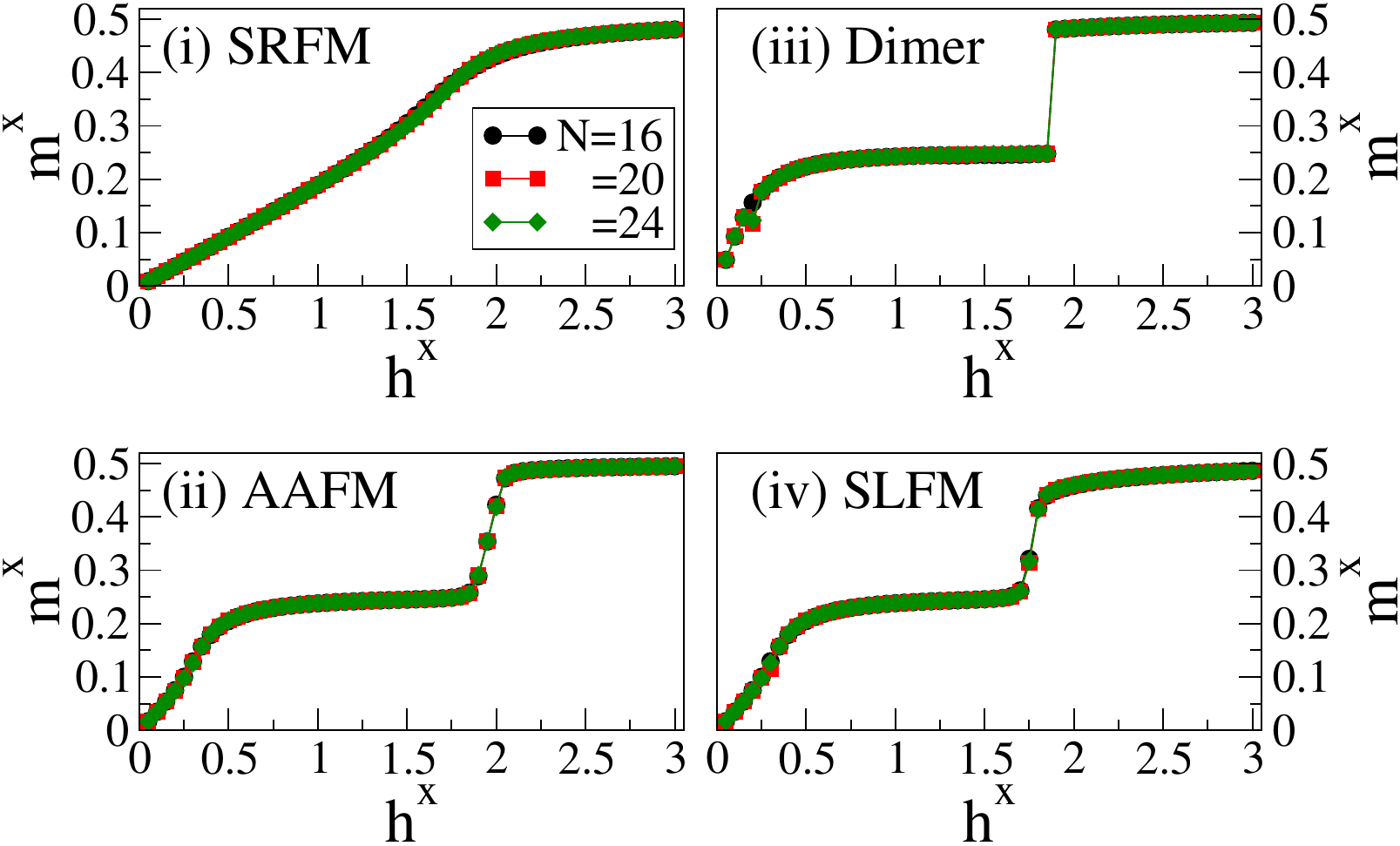}
\caption{Transverse  magnetization $m^x$ is shown as a function of the transverse field $h^x$ for four phases: (i) SRFM ($J_q=0.2$, $J_d=2.0$), (ii) AAFM ($J_q=2.0$, $J_d=0.4$) (iii) Dimer ($J_q=2.0$, $J_d=1.0$), and (iv) SLFM ($J_q=2.0$, $J_d=1.6$). Black, red, and green colors represent the system sizes $N=16, 20$, and $24$ respectively. }
\label{fig:magx}
\end{figure}
\begin{figure*}[t]
\includegraphics[width=1.0\linewidth]{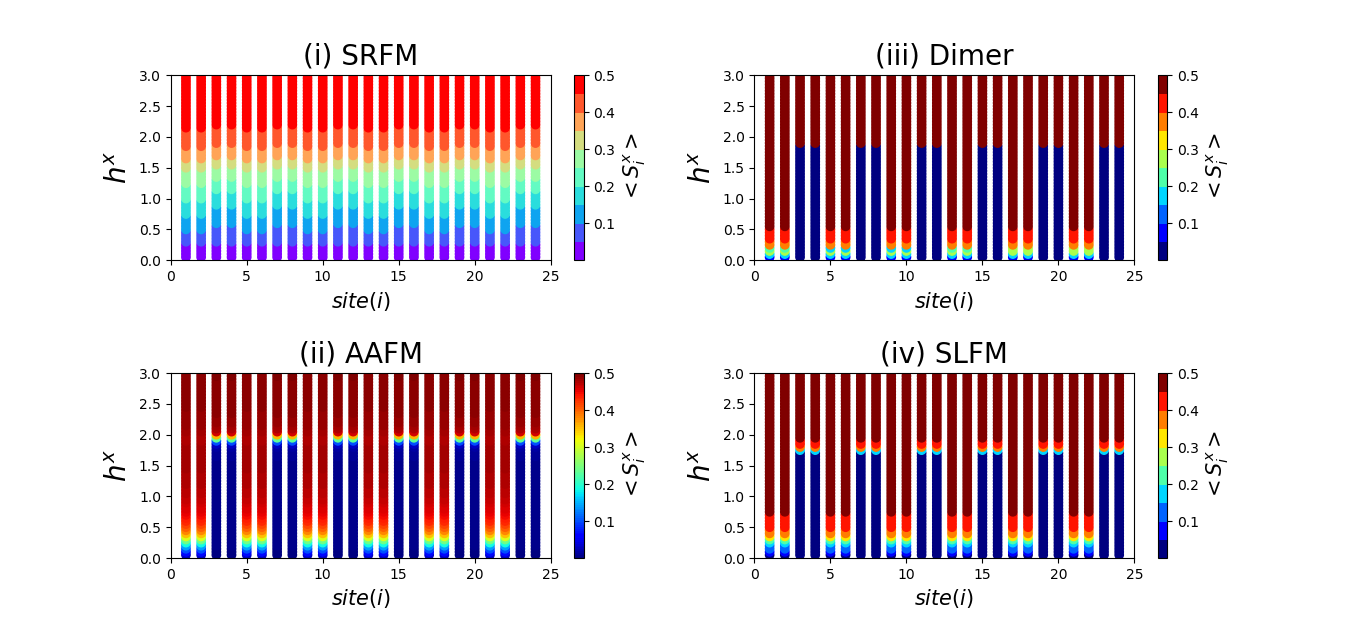}
\caption{Transverse component of spin density $\langle S^x_i\rangle$ at each site $i$ for for four phases: (i) SRFM ($J_q=0.2$, $J_d=2.0$), (ii) AAFM ($J_q=2.0$, $J_d=0.4$) (iii) Dimer ($J_q=2.0$, $J_d=1.0$), and (iv) SLFM ($J_q=2.0$, $J_d=1.6$) are shown for system size $N=24$. Along the horizontal axis, the site index $i$ for each spin is shown. Along the vertical axis, the transverse field $h^x$ is varied. The color bar shown in all subfigures represents the amplitude of $\langle S^x_i\rangle$ and it varies from $0$ to $0.5$ for spin-1/2 systems.}
\label{fig:spinx}
\end{figure*}
\subsubsection{Transverse component of magnetization}
It is to be mentioned that all four GS phases are in the $S^z=0$ sector, and so it does not change the longitudinal but rather the transverse component of magnetization on the application of an external transverse field $h^x$. We calculate the transverse magnetization $m^x$ in terms of spin density $\langle S^x_i \rangle$ at each site $i$ as- 
\begin{eqnarray}
\label{eqn:mx}
m^x=\frac{1}{N} \sum_{i=1}^{N} \langle S_i^x \rangle
\end{eqnarray}
In Fig.\ref{fig:magx}[(i)-(iv)], we show the transverse (along $+x$ direction) magnetization for all four phases: (i) the SRFM, (ii) the AAFM, (iii) the Dimer, and (iv) the SLFM with the corresponding set of chosen $J_q$, $J_d$ values as in section \ref{sec:magz} for the system sizes $N=16, 20$, and $24$. In the SRFM phase, the $m^x$ shows continuous variation with the field $h^x$ up to saturation value. In this phase, the GS has Ising bond dominance for all the spins, and therefore the spins along the x-direction get smoothly oriented along the field. In the AAFM phase, the curve increases smoothly and then it shows a $m^x \approx 1/4$ plateau-like behavior in the range $0.45<h^x<1.75$, and afterward it jumps to full saturation. A similar behavior is noticed in the Dimer phase as well in which the plateau onsets at a field $h^x=0.4$. In this case, the jump from the $m^x=1/4$ plateau to saturation is much faster than in the case of the AAFM phase. In the SLFM phase for $h^x<0.7$, $m^x$ increases smoothly, and then it forms the plateau-like structure for $0.7<h^x<1.7$. It shows a sudden jump almost around $h^x=1.75$ and slowly reaches saturation magnetization for higher $h^x$. From all the subfigures, it is noticed that there is a negligibly small finite-size effect. In the next subsection, we analyze the $m^x=1/4$ plateau mechanism for all the phases based on the spin density. 
\subsubsection{Transverse component of spin density}
For a more detailed understanding, we show the color map of the spin density $\langle S^x_i\rangle$ in all four phases for $N=24$. It is to be mentioned that the $\sigma-\sigma$ and $S-S$ pairs alternate with site index $i$. To be more specific, $\sigma-\sigma$ pairs take the site indices [{1,2},{5,6},....] whereas, $S-S$ pairs take the indices [{3,4},{7,8}....] as shown in all subfigures of Fig.\ref{fig:spinx}. In Fig.\ref{fig:spinx}.(i), $\langle S^x_i \rangle$ varies continuously as $h^x$ increases for all the sites. As the GS of the SRFM has only Ising interactions dominance and there is no transverse spin correlation, all the spins continuously get oriented along $h^x$ in this phase. In the AAFM phase, the $S-S$ pair has a strong transverse spin correlation whereas, the $\sigma-\sigma$ rung pairs are weak and are equations continuously with an increase in $h^x$ as shown in the color map Fig.\ref{fig:spinx}. (ii). Saturation is attained by an even further increase of field when it breaks the strong $S-S$ pair at $h^x=1.75$. As shown in Fig.\ref{fig:spinx}. (iii) for the Dimer phase, the continuous increase of $m^x$ is similar to the AAFM phase but a sudden jump at $h^x=1.7$ is noticed because of unbinding of the perfect $S-S$ singlet pairs. Fig.\ref{fig:spinx}.(iv) shows $\langle S^x_i\rangle$ for the SLFM phase. As in the GS of this phase, both the Ising pairs and Heisenberg pairs are aligned parallel but spins of opposite legs are aligned oppositely, with the increase in $h^x$, it is noticed that all the Ising dimer pairs are continuously broken until the magnetization reaches to $1/4$. An even further increase in $h^x$ does not easily close the spin gap and results in a large $m^x=1/4$ plateau until a field $h^x=1.75$ is applied to break the Heisenberg rung $S-S$ pairs.  
\section{Summary}
\label{sec:summary}
In this paper, we study the effect of external magnetic field on the GS phases of Hamiltonian in Eq.\ref{eqn:hamiltonian} of a frustrated spin-$\frac{1}{2}$ two-leg ladder with alternate Ising and Heisenberg type of rung and Ising type of interaction in the leg and diagonal. This model has a few similarities to one of the earlier studied models where leg and diagonal exchanges are Ising type but all the rung exchanges are Heisenberg type in which on applying the magnetic field, a magnetic phase SB appears \cite{verkholayak2012}. In our model, alternate rungs are Ising-Heisenberg type due to which the zero field GS phases and the plateau phases depend upon the competing Ising and Heisenberg rung exchanges as well. Tuning of the exchange parameters in the model Hamiltonian can give rise to four GS phases: (i) the SRFM, (ii) the AAFM, (iii) the Dimer, and (iv) the SLFM, whose spin arrangements are schematically represented in Fig.\ref{fig:model2}.b. [(i)-(iv)]. We analyzed the magnetization behavior in the presence of external magnetic fields: longitudinal and transverse. 

In the presence of the longitudinal magnetic field $h$, the GS shows three magnetic plateaus: the first one is due to the finite spin gap, the second plateau at $m=1/4$ is formed due to the polarization of either Ising or Heisenberg type of rung spin dimers along the field. This can be of two types as shown in Fig.\ref{fig:configuration}.b. In the SRFM phase, the Heisenberg rung spin pairs $S-S$ are polarized giving rise to ``$m=1/4$ plateau type-1'' or ``P-I" phase  as shown in Fig.\ref{fig:configuration}.b.(i). But for the other three zero field GS phases: the AAFM, the Dimer, and the SLFM, the $\sigma-\sigma$ spins are polarized at $m=1/4$ and give rise to ``$m=1/4$ plateau type-2" or ``P-II" phase as shown in Fig.\ref{fig:configuration}.b.(ii). In the presence of a large external field in all phases, all of the spins are completely polarized along the field and form the third plateau at the saturation magnetization $m=1/2$ which is named FP. The $m=1/4$ plateau width is sensitive to the parameter values as obtained in Eq.\ref{eqn:plateauwidth}. We also notice that two plateaus are connected by jumps in the magnetization curve and this is because of the unpairing of either all $S-S$ (for P-I) or all $\sigma-\sigma$ (for P-II) rung dimers.

To understand the quantum nature of the wave function of the GS, we calculate the quantum fidelity and quantum concurrence in the presence of a longitudinal field. In all four phases, fidelity shows deviation from unity at the critical fields of magnetic phase transitions as shown in Fig.\ref{fig:fflvsh}. We note that the discontinuities in fidelity represent the overlap of the plateau phases. The quantum concurrence shown in Fig.\ref{fig:concvsh}.c measures the entanglement between two spins at the Heisenberg rung. The concurrence is always zero as a function of the field for the SRFM phase, and it means that the SRFM phase is a pure state. Whereas, in other phases: the AAFM, the Dimer, and the SLFM, the concurrence has a finite value both at $m=0$. In the $m=0$ plateau phase, the concurrence is a function of the exchange interactions for the AAFM, and the SLFM phases and it is maximum for the Dimer phase. The concurrence is zero for the P-I phase and FP phase. Whereas it is maximum for the P-II phase as it forms the isolated singlet Dimers along the Heisenberg rung. So, one can state that the P-II phase is maximally entangled, whereas, P-I and FP phases are the pure states. However, all of the jumps in the magnetization can be indirectly predicted based on the jumps in concurrence as shown in Fig.\ref{fig:concvsh}.c.

We define QPI to uniquely identify the quantum phases in the exchange parameter $J_q$-$J_d$ plane for a finite longitudinal field $h$. The unique QPI values for different quantum phases are tabulated in Table.\ref{tab:table}. The effect of the longitudinal magnetic field on the GS phases is shown in Fig.\ref{fig:magnetic_phases}.[(i)-(iv)]. For a sufficiently larger longitudinal field $h=1.0$, the Dimer phase disappears, and two types of $m=1/4$ plateaus: P-I, P-II appear. At a very large field $h=3.0$, it is noticed that the P-I, P-II, and FP phases dominate in the phase diagram. The phase boundaries among the quantum phases can easily be obtained from the analytical expression of the critical fields of phase transitions.

In the SRFM phase, spin alignments are along the z direction and there is weak exchange interaction along the +x direction in the Heisenberg rung dimers, therefore, magnetization $m^x$ shows a continuous variation till saturation on the application of a transverse field. In other phases: the AAFM, the Dimer, and the SLFM, the magnetization process can be understood in terms of two sublattice behaviors. The sublattice with $\sigma-\sigma$ dimer is paramagnetic along x, whereas, in the other sublattice, the Heisenberg spin pair $S-S$ dimer has a strong transverse exchange component which induces a finite spin gap in the system. As a consequence, at a lower value of the transverse field, the system shows a continuous behavior in the magnetization curve due to gradual change of $m^x$ of $\sigma-\sigma$ spins up to $m^x=1/4$ with an increase in $h^x$  as shown using spin density in Fig.\ref{fig:spinx}.[(ii)-(iv)]. With further increase in field, at a critical value, the magnetization curve shows a sudden jump from $m^x \approx 1/4$ to $1/2$ for the three phases: the AAFM, the Dimer, and the SLFM, and this phase transition seems to be of the first order. The plateau width is sensitive to the set of parameters or the set of exchange interactions $J_q$ and $J_d$ in the presence of a transverse field also. 

In conclusion, this model system is unique because of the alternate Ising-Heisenberg rung exchange and gives many insightful mechanisms of the plateau and jumps both in the presence of longitudinal and transverse fields. The magnetic properties of such systems can be utilized in designing quantum switches, magnetic memories, and other similar devices. Also, these systems might have tremendous applications in quantum information processing and quantum computation because of the entangled states.

\section{Acknowledgements}
M.K. thanks SERB Grant Sanction No. CRG/2020/000754 for the computational facility. S.S.R. thanks CSIR-UGC for financial support.

\begin{strip}
\section{Appendix 1}
\label{appendix}
         The partition function in presence of a longitudinal field $h$ for $N$ sites, $Q_N(h, \beta)$ with Hamiltonian \textbf{$H$} can be written as-
\begin{eqnarray}\label{eq:def_partition}
\centering
      &&  Q_{N}(h,\beta)= Tr \left( e^{-\beta \textbf{H}} \right)  %\nonumber
\end{eqnarray}
       Where, Tr means trace of the matrix, $\beta= 1/\left(k_{B}T\right)$ and $k_{B}$
       is the Boltzmann constant.
       Using explicit configuration basis for the system, Eq. \ref{eq:def_partition} is rewritten in the following form,
 \begin{eqnarray}%
 \label{eq:partition_basis0}
      &&Q_{N}(h,\beta)=\sum_{\{\sigma,S\}} 
     <\cdots, \sigma_{2j-1,1},\sigma_{2j-1,2},S_{2j,1},S_{2j,2},\cdots \mid 
         e^{-\beta \mathbf{H}} \mid        
     \cdots, \sigma_{2j-1,1},\sigma_{2j-1,2},S_{2j,1},S_{2j,2},\cdots >. \nonumber 
 \end{eqnarray}
Here the summation is over all possible configurations $\{\sigma,S\}$ of the system. For a given configuration,
$|\cdots, \sigma_{2j-1,1},\sigma_{2j-1,2},S_{2j,1},S_{2j,2},\cdots >$ represents a basis state. In our case, the system is composed of $n=N/4$ units, and for each unit, the Hamiltonian is written in Eq.\ref{eqn:hamiltonian}.  The partition function of the entire ladder can be written as:
\begin{eqnarray}%\label{eq:partition_basis1}
      && Q_{N}(h,\beta)= \sum_{\sigma} <\cdots,\sigma_{2j-1,1},\sigma_{2j-1,2},\cdots \mid \prod_{j=1}^{n} 
                   \mathbf{T}_j \mid \cdots,\sigma_{2j-1,1},\sigma_{2j-1,2},\cdots> \nonumber
\end{eqnarray}
where $T_i=\sum_{\{S\}_i}<S_{2i,1},S_{2i,2}\mid e^{-\beta \mathbf{H_{i}}(\sigma,S)}\mid S_{2i,1},S_{2i,2}>$ is well-known transfer matrix operator for each unit. Here the
summation is over $\{S\}_i$ which represents all possible configurations of spins $S_{2i,1}$ and $S_{2i,2}$ (from the $i^{th}$ unit).
It may be noted
that  $T_i$ does not contain the components of spin $S$ operators and it has only $\sigma$
variables, namely, $\sigma_{2i-1,1},\sigma_{2i-1,2}, \sigma_{2i+1,1}$ and $\sigma_{2i+1,2}$.
Since, the Hamiltonians of each unit commute to each other, introducing identity operators
$I = \sum_{\{\sigma\}_i}|\sigma_{2i-1,1},\sigma_{2i-1,2}> <\sigma_{2i-1,1},\sigma_{2i-1,2}|$ between successive $\mathbf{T}$ operators, we can finally write the
partition function as the trace of the $n$-th power of a small ($4\times4$) transfer matrix $\mathbf{P}$. We have $Q_{N}(h,\beta)= Tr ( \mathbf{P}^n)$. The elements of the transfer matrix are given by
%\begin{widetext}
\begin{eqnarray}\label{eq:pmat_elements}
      &&  P_{(\sigma_{2i-1,1},\sigma_{2i-1,2}),(\sigma_{2i+1,1},\sigma_{2i+1,2})} = 
<\sigma_{2i-1,1},\sigma_{2i-1,2}\mid \mathbf{T}_i \mid \sigma_{2i+1,1},\sigma_{2i+1,2}>
\end{eqnarray}
%\end{widetext}
Before we construct and diagonalize the $\mathbf{P}$ matrix, we first need to carry out the trace over the configurations $\{S\}_i$ to find out the form of
$\mathbf{T}_i$. Since $\mathbf{T}_i=\sum_{\{S\}_i}<S_{2i,1},S_{2i,2}\mid e^{-\beta \mathbf{H_{i}}(\sigma,S)}\mid S_{2i,1},S_{2i,2}>$,
if we take the eigenstate
basis of $\mathbf{H}_i$, we will get $\mathbf{T}_i$ as the summation over exponential of eigenvalues of $-\beta\mathbf{H}_i$. Next, we calculate the eigenvalues of
$\mathbf{H}_i$ operator. Now, y considering, \\
      $  a=J_{d} \left( \sigma_{2j-1,2}^{z}+ \sigma_{2j+1,2}^{z} \right) +J_{cq}\left( \sigma_{2j-1,1}^{z}+
          \sigma_{2j+1,1}^{z}\right)$
       +$h $  \\
      $  b=J_{d} \left( \sigma_{2j-1,1}^{z}+ \sigma_{2j+1,1}^{z} \right) +J_{cq}\left( \sigma_{2j-1,2}^{z} +
          \sigma_{2j+1,2}^{z}\right)$
        +$h $  \\
      $  c=\frac{J_{c}}{2}\left(\sigma_{2j-1,1}^{z}\sigma_{2j-1,2}^{z} +
          \sigma_{2j+1,1}^{z}\sigma_{2j+1,2}^{z}\right) $ \\
      $  d=\frac{h}{2}\left(\sigma_{2j-1,1}^{z}+\sigma_{2j-1,2}^{z}+\sigma_{2j+1,1}^{z}+\sigma_{2j+1,2}^{z}\right) $, \\
      $f=c+d$ \\
         Hamiltonian (Eq. \ref{eqn:hamiltonian} ) for the $j^{th}$ geometrical unit can be written as-
\begin{eqnarray} \label{eq:operators}
\centering
      &&  \mathbf{H_{j}}= \frac {J_q^{xy}}{2}\left(S_{2j,1}^{+}S_{2j,2}^{-}+S_{2j,1}^{-}S_{2j,2}^{+}\right)+
           J_q^{z}\left(S_{2j,1}^{z}S_{2j,2}^{z}\right)   
        +a S_{2j,1}^{z}+b S_{2j,2}^{z}+f           \nonumber
\end{eqnarray}
       We can write down the following Hamiltonian matrix in the eigenstate basis of   $S_{2j,1}^{z}S_{2j,2}^{z}$ operator as \\
\[ H_j = \left[ \begin{array}{cccc}
\frac{J_q^{z}}{4}+\frac{(a+b)}{2}+f & 0 & 0 & 0 \\
0 & \frac{-J_q^{z}}{4}+\frac{(a-b)}{2}+f & \frac{J_q^{xy}}{2} & 0 \\
0 & \frac{J_q^{xy}}{2} & \frac{-J_q^{z}}{4}-\frac{(a-b)}{2}+f & 0 \\
0 & 0 & 0 & \frac{J_q^{z}}{4}-\frac{(a+b)}{2}+f \\
\end{array}
\right]\]
\end{strip}
%\end{widetext}

         The Hamiltonian matrix comes up with its four eigenvalues from three $S^z_{SS}$ sectors based on S-S pairs- \\
\textbf{(i) From $S^z_{SS}=1$ sector (formed by S-S pair)} \\
       $ \theta_1=(f+\frac{J_q^{z}}{4})+\frac{(a+b)}{2} $ \\
\textbf{(ii) From $S^z_{SS}=-1$ sector (formed by S-S pair)} \\
       $ \theta_2=(f+\frac{J_q^{z}}{4})-\frac{(a+b)}{2} $ \\
\textbf{(iii) From $S^z_{SS}=0$ sector (formed by S-S pair)} \\
       $ \theta_3=(f-\frac{J_q^{z}}{4}) + \frac{\sqrt{(J_q^{xy})^2+(a-b)^2}}{2} $ \\
       $ \theta_4=(f-\frac{J_q^{z}}{4}) - \frac{\sqrt{(J_q^{xy})^2+(a-b)^2}}{2} $. \\
We note that the eigenvalues ($\theta_{k'}$) are functions of $\sigma$ variables, namely $\sigma_{2j-1,1},\sigma_{2j-1,2}, \sigma_{2j+1,1}$ and $\sigma_{2j+1,2}$.
Using these eigenvalues, we rewrite $\mathbf{T}_j$ as,
\begin{eqnarray}
&\mathbf{T}_j = \sum_{\{S\}_j}<S_{2j,1},S_{2j,2}\mid e^{-\beta \mathbf{H_{j}}(\sigma,S)}\mid S_{2j,1},S_{2j,2}> \nonumber \\
&~~~= \sum_{k'=1}^4 e^{-\beta \theta_k'}.\nonumber
\end{eqnarray}

\begin{eqnarray}
& =2e^{-\beta f}\bigg[e^{-\frac{\beta J_q^{z}}{4}} Cosh\bigg(\frac{\beta (a+b)}{2}\bigg)  \nonumber \\
& +e^{\frac{\beta J_q^{z}}{4}} Cosh\bigg( \frac{\beta J_q^{xy}}{2} \sqrt{1+\frac{(a-b)^2}{(J_q^{xy})^2}}\bigg)\bigg] \nonumber
\end{eqnarray}

Further, we consider- $e^{\frac{\beta J_q^{z}}{4}}=Q$ \space,  $e^{\frac{\beta J_c}{4}}=C$, \\
$e^{\frac{\beta h}{4}}=H$, \space $\frac{(J_{cq}+J_d)}{2}=X$, 
\space $\frac{(J_{cq}-J_d)^2}{(J_q^{xy})^2}=Y$, \\
and also, $\Delta_1=\sqrt{1+Y}$, \space
$\Delta_2=\sqrt{1+4Y}$.   \\
The Transfer matrix for one unit becomes -
\[ \mathbf{P} = \left[ \begin{array}{cccc}
p       & q     & q         & r             \\
q       & s     & u         & v             \\
q       & u     & s         & v             \\
r       & v     & v         & w              \\
\end{array}
\right].\]
Where,
\begin{eqnarray}
&p=2e^{-\beta(J_c/4+h)}\nonumber \\ 
&\times \bigg[
Q^{-1} Cosh \big( \beta(2X+h) \big)+
Q  Cosh \big(\frac{\beta J_q^{xy}}{2} \big)
\bigg] \nonumber
\end{eqnarray}
\begin{eqnarray} 
&q=2e^{-\beta(h/2)} \nonumber \\ 
&\times \bigg[
Q^{-1} Cosh \big( \beta(X+h) \big)+
Q  Cosh \big(\frac{\beta J_q^{xy} \Delta_1}{2} \big)
\bigg] \nonumber
\end{eqnarray}
\begin{eqnarray}
&r=2e^{-\beta(J_c/4)}\nonumber \\ 
&\times \bigg[
Q^{-1}Cosh \big( \beta h \big)+
Q  Cosh \big( \frac{\beta J_q^{xy}}{2} \big)
\bigg] \nonumber
\end{eqnarray}
\begin{eqnarray}
&s=2e^{\beta(J_c/4)} \nonumber \\
& \times \bigg[
Q^{-1}Cosh \big( \beta h \big)+
Q  Cosh \big(\frac{\beta J_q^{xy} \Delta_2}{2} \big)
\bigg] \nonumber
\end{eqnarray}
\begin{eqnarray}
&u=2e^{\beta(J_c/4)} \nonumber \\
& \times \bigg[
Q^{-1} Cosh \big( \beta h \big)+
Q  Cosh \big(\frac{\beta J_q^{xy}}{2} \big)
\bigg] \nonumber 
\end{eqnarray}
\begin{eqnarray}
&v=2e^{\beta(h/2)} \nonumber \\
&\times \bigg[
Q^{-1} Cosh \big( \beta (-X+h) \big)+
Q  Cosh \big( \frac{\beta J_q^{xy} \Delta_1}{2} \big)
\bigg] \nonumber
\end{eqnarray}
\begin{eqnarray}
&w=2e^{-\beta(J_c/4-h)} \nonumber \\ 
& \times \bigg[
Q^{-1} Cosh \big( \beta (-2X+h) \big)+
Q Cosh \big( \frac{\beta J_q^{xy}}{2} \big)
\bigg] \nonumber 
\end{eqnarray}

 From $|P-\lambda I_4|=0$, we get the eigenvalues in the form of
\begin{eqnarray}
\label{eq:pol3}
& \fbox{$\lambda_4=(s-u)$} \nonumber \\
& \fbox{$\lambda^3-B_0 \lambda^2 -C_0 \lambda +D_0 =0$}
\end{eqnarray}
Here, $\lambda_4$ is one of the eigenvalues, whereas, the other three come from Eq.\ref{eq:pol3}. The coefficients of the equation are defined as:
\begin{eqnarray}
& B_0=(s+u+w) \nonumber \\
& C_0=[2(q^2+v^2-\frac{r^2}{2})-pw+(p-w)(s+u)] \nonumber \\
& D_0=[4qrv-2pv^2-2q^2w+pw^2-(s+u)(r^2+pw)] \nonumber
\end{eqnarray}
For the polynomial equation \ref{eq:pol3}, the eigenvalues $\lambda_i$ satisfy the relations-
\begin{eqnarray}
\label{eqn:relationcoeff1}
\sum^3_{i=1} \lambda_i=B_0,\sum^3_{i=1} \lambda_i \lambda_{i+1}=-C_0
\end{eqnarray}
Now, let us make a very reasonable assumption to make the calculation easy. We assume, $\lambda_1 \gg \lambda_2 \gg \lambda_3$ are in descending order and $\lambda_3$ has the least contribution in the partition function so that Eq.\ref{eqn:relationcoeff1} can approximately be written as:
\begin{eqnarray}
\label{eqn:relationcoeff2}
 \lambda_1+\lambda_2=B_0,
 \lambda_1 \lambda_2=-C_0
\end{eqnarray}
The Eq. \ref{eqn:relationcoeff2} leads us to getting other two eigenvalues:
\begin{eqnarray}
\label{eq:largeeigen}
& \lambda^2_{1,2}-B_0 \lambda_{1,2}-C_0=0 \nonumber \\
& \fbox{$\Longrightarrow   \lambda_{1,2} = \frac{B_0 \pm \sqrt{B^2_0+4C_0} }{2}$}
\end{eqnarray}
We find Eq. \ref{eq:largeeigen} becomes much more simpler with further approximation in $\beta\rightarrow\infty$ limit as-
\begin{eqnarray} 
\label{eqn:approx_Q4}
\centering
& \lambda_1=(w+s+u) \nonumber \\
& =  2e^{\frac{-\beta (J_c-4h)}{4}}\bigg[
Q^{-1}Cosh[\beta (h-2X)] + Q Cosh[\frac{\beta J_q^{xy}}{2}]
\bigg] \nonumber \\
& +\bigg[2 Q^{-1}Cosh[\beta h]+Q Cosh[\frac{\beta J_q^{xy}}{2}]+Q Cosh[\frac{\beta J_q^{xy}\Delta_2}{2}]
\bigg] \nonumber \\
& \times 2e^{\frac{\beta (J_c)}{4}} 
\end{eqnarray}
$Q_N(h,\beta)$ takes the form as \\
$Q_N(h,\beta)=[\lambda_1^n+\lambda_2^n+\lambda_3^n+\lambda_4^n]$ \\
$\rightarrow$ $Q_N(h,\beta)=\lambda_1^n[1+\frac{\lambda_2}{\lambda_1}^n+\frac{\lambda_3}{\lambda_1}^n+\frac{\lambda_4}{\lambda_1}^n]$ \\
For $n \rightarrow \infty$, and $\lambda_1$ being the largest, the partition function for the entire system and one unit become $Q_N(h,\beta) \approx \lambda_1^n$ and $Q_4(h,\beta) \approx \lambda_1$ respectively.
%%%%%%%%%%%%%%%%%%%%%%%%%%%%%%%%%%%%%%%%%%%%%%%%%%%%%%%%%%%%%%%%%%%%%%%%%%%%% 
%%%%%%%%%%%%%%%%%%%%%%%%%%%%%%%%%%%%%%%%%%%%%%%%%%%%%%%%%%%%%%%%%%

%																 %
%																 %
%		REFERENCES						o					     %
%																 %
%																 %
%%%%%%%%%%%%%%%%%%%%%%%%%%%%%%%%%%%%%%%%%%%%%%%%%%%%%%%%%%%%%%%%%%
\section*{References}
\bibliographystyle{iopart-num}
\bibliography{ref_ladder}
\end{document}